\documentclass[%
preprint,
 amsmath,amssymb,
aps, showkeys
]{revtex4-2}

\usepackage{graphicx}% Include figure files
\usepackage{xcolor}
\usepackage{dcolumn}% Align table columns on decimal point
\usepackage{bm}% bold math
\usepackage{placeins}
\usepackage{mathrsfs}  
\usepackage{amsmath}
\usepackage{amssymb}
\usepackage{amsfonts}
\usepackage{dsfont}

\usepackage{tabularx}
\usepackage{graphicx}
\usepackage{float}
\usepackage{caption}
\usepackage{subcaption}
\usepackage{framed}

\usepackage{comment}
\usepackage{appendix}
\usepackage{tcolorbox}
\usepackage[colorinlistoftodos]{todonotes}

\usepackage{enumitem}
\usepackage{cleveref}

\newcommand*\diff {\mathop{}\!\mathrm{d}}

\newcommand{\mean}[1] {
\ensuremath{\left\langle #1 \right\rangle}}

\begin{document}

\newcommand\blfootnote[1]{%
  \begingroup
  \renewcommand\thefootnote{}\footnote{#1}%
  \addtocounter{footnote}{-1}%
  \endgroup
}

\preprint{APS/123-QED}

\title{Revisiting the framework for intermittency in Lagrangian stochastic models for turbulent flows: a way to an original and versatile numerical approach}
%\title{A revisited framework for intermittency in Lagrangian stochastic models for turbulent flows paving the way to an original and versatile numerical approach}

\author{Roxane Letournel}
 \email{roxane.letournel@centralesupelec.fr}
 \affiliation{
 Laboratoire EM2C, CNRS, CentraleSup\'elec, Universit\'e Paris-Saclay, 3 rue Joliot Curie, 91192 Gif-sur-Yvette cedex, France
 }
 \affiliation{
F\'ed\'eration de Math\'ematiques de CentraleSup\'elec, CNRS FR-3487, CentraleSup\'elec, Universit\'e Paris-Saclay, 9 rue Joliot Curie, 91190 Gif-sur-Yvette cedex, France
}
\affiliation{
CMAP, CNRS, \'Ecole polytechnique, Institut Polytechnique de Paris, Route de Saclay, 91128 Palaiseau cedex, France
}

\author{Ludovic Gouden\`ege}
 \affiliation{
F\'ed\'eration de Math\'ematiques de CentraleSup\'elec, CNRS FR-3487, CentraleSup\'elec, Universit\'e Paris-Saclay, 9 rue Joliot Curie, 91190 Gif-sur-Yvette cedex, France
}

\author{R\'emi Zamansky}
 \affiliation{
Institut de M\'ecanique des Fluides de Toulouse (IMFT), Universit\'e de Toulouse, CNRS-INPT-UPS, Toulouse FRANCE
 }
 
 \author{Aymeric Vi\'e}
 \affiliation{
 Laboratoire EM2C, CNRS, CentraleSup\'elec, Universit\'e Paris-Saclay, 3 rue Joliot Curie, 91192 Gif-sur-Yvette cedex, France
 }

\author{Marc Massot}
\affiliation{
CMAP, CNRS, \'Ecole polytechnique, Institut Polytechnique de Paris, Route de Saclay, 91128 Palaiseau cedex, France
}

\date{\today}% It is always \today, today,
             
%\blfootnote{Preprint submitted to Physical Review E. March, 2021}
\begin{abstract}

The characterization of intermittency in turbulence has its roots in the K62 theory, and if no proper definition is to be found in the literature, statistical properties of intermittency were studied and models were developed in attempt to reproduce it. 
The first contribution of this work is to propose a requirement list to be satisfied by models designed within the Lagrangian framework. 
Multifractal stochastic processes are a natural choice to retrieve multifractal properties of the dissipation. 
Among them, following the proposition of \cite{Mandelbrot1968}, we investigate the Gaussian Multiplicative Chaos formalism, which requires the construction of a log-correlated stochastic process $X_t$. 
The fractional Gaussian noise of Hurst parameter $H = 0$ is of great interest because it leads to a log-correlation for the logarithm of the process.
Inspired by the approximation of fractional Brownian motion by an infinite weighted sum of correlated Ornstein-Uhlenbeck processes, our second contribution is to propose a new stochastic model: $X_t = \int_0^\infty Y_t^x k(x) d x$, where $Y_t^x$ is an Ornstein-Uhlenbeck process with speed of mean reversion $x$ and $k$ is a kernel. A regularization of $k(x)$ is required to ensure stationarity, finite variance and logarithmic auto-correlation. A variety of regularizations are conceivable, and we show that they lead to the aforementioned multifractal models.
To simulate the process, we eventually design a new approach relying on a limited number of modes for approximating the integral through a quadrature $X_t^N = \sum_{i=1}^N \omega_i Y_t^{x_i}$, using a conventional quadrature method. This method can retrieve the expected behavior with only one mode per decade, making this strategy versatile and computationally attractive for simulating such processes, while remaining within the proposed framework for a proper description of intermittency.
\end{abstract}

%Use showkeys class option if keyword
\maketitle

%\tableofcontents
\vspace*{2cm}

\section{Introduction}
\label{sec:Introduction}
%\textbf{Why is it important to retrieve intermittency ?}\\
The stochastic nature of turbulence and the statistical behaviors of velocity fluctuations have been widely investigated, 
in order to understand and then reproduce its properties on reduced turbulence models (Large Eddy Simulation) (see \cite{Minier2014}). 
The inertial scales of many turbulent flows are correctly described by the classical image of Richardson's energy cascade \cite{Richardson1922}. 
Kolmogorov formalized this universality of turbulence with a self-similar description of velocity fluctuations in the inertial range (see \cite{Kolmogorov1941}, hereafter referred as K41). 
However, it was  pointed out in Ref.~\cite{Landau1944} that this theory is flawed at small scales by the phenomenon of intermittent energy dissipation, in contradiction with the homogeneity assumed in K41. \\

 Kolmogorov and Obukhov developed, in response to that concern, a vision based on local and scale-dependent observables which are more relevant to describe velocity fluctuations (see \cite{Kolmogorov1962}, hereafter referred as K62). Since the publication of the refined similarity hypotheses, many studies have been devoted to data analysis, most of them focusing on energy dissipation. 
 Consistently with these hypotheses, it was observed that the dissipation has a log-normal distribution and presents long-range power-law correlation (see Refs.~\cite{Yeung1989, Pope1990, Yeung2006a, Dubrulle2019}). Reproducing such behaviors in turbulence simulations is still an open problem and we are interested in the derivation of models that retrieve this intermittency in Reynolds Averaged Navier-Stokes (RANS) modelings or Large Eddy Simulation (LES) contexts, in particular for modeling phenomena related to the small scales, such as combustion instabilities or the atomization of droplets in industrial burners...\\

%\textbf{Multifractal random fields} \\
Multifractal random fields are of primary interest for modeling intermittent fields since  they possess high variability on a wide range of time or space scales, associated with intermittent fluctuations and long-range power-law correlations \cite{Borgas1993,Frisch1995,Sreenivasan1997}. As opposed to monofractal, self-similar fields that correspond to the K41 description of turbulence, complex structures observed in DNS and experimental studies are well reproduced by multifractal random fields. \\

%\textbf{From discrete to continuous models}\\
The multiplicative cascade model of Yaglom  \cite{Yaglom1966} is at the basis of most cascade models introduced later to account for turbulent intermittency. 
It was able to reproduce both experimental facts and Kolmogorov's log-normal hypothesis. 
Discrete models picture turbulence as an ensemble of discrete length-scales, in which the energy transfers from a “mother" to a “daughter" eddy in a recursive and multiplicative manner. In this way, large fluctuations recursively generate correlations over long distances.
 Other discrete models were also formulated later and the reader is referred to the exhaustive review of \cite{Seuront2005}.
However, in \cite{Mandelbrot1968}, Mandelbrot criticized these models for being based on a discrete and arbitrary ratio of length scales. They suggested to consider continuous models such as Gaussian Multiplicative Chaos which was later formalized in Refs.~\cite{Kahane1985, Robert2010}. The second criticism of \cite{Mandelbrot1968} concerns the early cascade models that were developed in the Eulerian framework, and therefore do not exhibit a spatio-temporal structure, which was then introduced by mean of stochastic causal models. The Lagrangian framework of intermittency was proposed in \cite{Borgas1993} and equivalent behavior of multifractal properties were observed for dissipation along particle trajectories. 
Causal and sequential multifractal stochastic processes were developed in response (see Refs.~\cite{Biferale1998, Schmitt2001, Muzy2002, Chevillard2017a, Pereira2018}). However, most of them rely on long-term memory stochastic processes and can be computationally expensive. \\

%\textbf{This work}\\
The first objective of this work is to establish a list of criteria for modeling intermittent dissipation. This characterization is based on observations of experimental data and is in accordance with the phenomenology developed by Kolmogorov. We show to what extent the  Gaussian Multiplicative Chaos formalism is relevant for the proposed requirements. \\
Secondly, this work aims at developing a general framework for causal stochastic models based on the Gaussian Multiplicative Chaos. In particular, the novelty lies in the construction of a log-correlated stochastic process $X_t^\infty$.
Introducing the inverse Laplace transform of kernel functions of fractional Brownian motions \cite{Mandelbrot1968}, it is possible to express such stochastic processes by mean of an infinite sum of Ornstein-Uhlenbeck processes. Such formulation is discussed and regularizations are proposed to ensure multifractal properties of the stochastic process $X_t^\infty$ in the inertial range. 
We eventually show that the newly introduced formulation encompasses most of the existing models. \\
Finally, we develop a numerical method for simulating this process, based on a quadrature of the infinite sum, i.e. on a finite sum of Ornstein-Uhlenbeck processes. This method is a discrete version of the process $X_t^\infty$ and
has the benefit of being computationally affordable and versatile. 
This discretization can be seen as the selection of representative time-scales for the few Ornstein-Uhlenbeck processes all along the inertial range. The densification of these time-scales corresponds to the continuous model $X_t^\infty$: with an infinity of time-scales, each one assigned to a turbulent structure and thus offers a natural physical interpretation. \\
Let us underline that stochastic calculus plays a crucial role in introducing and analyzing this modeling process as well as the asymptotic and singular limits. The purpose and scope of the present paper is
related to the physical relevance of the introduced concepts and their impact in terms of numerical simulations of intermittent turbulent flows. Since the mathematical foundations of the results we use are out of the scope of the paper, we 
refer to a companion paper \cite{Goudenege2021}, where we propose a synthesis of the mathematical key results and their justification in terms of stochastic calculus. \\

%\textbf{Organization of the paper}\\
The paper is organized as follows: in section \ref{sec:Multifractal_intermittency} we discuss the origins and the properties of intermittent dissipation for turbulent flows and we provide a characterization of it. We also recall the Gaussian Multiplicative Chaos formalism. 
Section \ref{sec:Infinite_sum_OU} presents the procedure to express any fractional Brownian motion by an infinite sum of Ornstein-Uhlenbeck processes. Inspired by this formulation, we examine this new process and introduce necessary conditions to ensure its intermittency. We point out that this general formulation encompasses previous causal stochastic models. 
Then, in section~\ref{sec:Infinite_sum_OU}, the numerical procedure to simulate the proposed stochastic process is described and we discuss the benefits and the physical grounds of such modeling. 

\section{Properties of intermittency in turbulence and multifractal models}
\label{sec:Multifractal_intermittency}
\subsection{Origins and properties of the intermittency}
\label{subsec:Characterization_Intermittency}
In Ref.~\cite{Kolmogorov1941}, Kolmogorov first formalized the vision of Richardson cascade: “Big whirls have little whirls that feed on their velocity, and little whirls have lesser whirls and so on to viscosity" by introducing the “Similarity hypothesis". He stated that for high Reynolds numbers $\mathrm{Re}_L = \dfrac{\sigma_u L}{\nu}$, where $\sigma_u$ is the velocity standard deviation, $L$ the characteristic length-scale of fluid stirring, and $\nu$ the viscosity, turbulence is universal and velocity fluctuations statistics are expected to be independent of the large scales. 
Based on these two parameters, Kolmogorov scales can be introduced: $\eta \equiv \left( \nu^3 / \mean{\varepsilon} \right)^{1/4}$, $u_\eta \equiv \left(  \mean{\varepsilon} \nu \right)^{1/4}$, $\tau_\eta \equiv \left( \nu / \mean{\varepsilon} \right)^{1/2}$.  \\ 
The K41 states that in the inertial range, there is a complete similarity, and velocity increments statistics along Lagrangian trajectories are independent on viscosity and therefore only determined by the mean dissipation $\mean{\varepsilon}$ and the time scale $\tau$:
\begin{equation}
\mean{[\Delta_\tau u]^2} = C_0 \mean{\varepsilon} \tau, \quad \text{for } \tau_\eta \ll \tau \ll T_L
\end{equation}
 where $\Delta_\tau u = u(t+\tau) - u(t)$ is the velocity increment along a fluid particle trajectory and $C_0$ the universal Lagrangian velocity structure function constant. The inertial range lies from the Kolmogorov time scale $\tau_\eta$ and the integral time scale $T_L = (1/\sigma_u^2) \int_0^\infty \mean{u(t)u(t+\tau)}\diff \tau$ which is the characteristic time of correlation of fluid particle velocity. 
Similar arguments with Eulerian velocity structure functions, defined with space increments, lead to the well-known theoretical “-5/3" power law of the spatial energy Spectrum.\\ 
The conservation of the rate of energy transfer in the inertial range, given by the Kolmogorov scaling $\mean{\varepsilon} \sim \dfrac{\sigma_u^2}{T_L} \sim \dfrac{\sigma_u^3}{L} \sim \dfrac{u_\eta^2}{\tau_\eta} $, allows us to obtain:
\begin{equation}
\dfrac{L}{\eta} \sim \mathrm{Re}_L^{3/4}, \quad
\dfrac{T_L}{\tau_\eta} \sim \mathrm{Re}_L^{1/2} 
\end{equation}
%The Reynolds number based on Taylor microscale $\lambda$ is directly proportional to the ratio of the timescales : $\mathrm{Re}_\lambda \sim \dfrac{T_L}{\tau_\eta}$. \\ 

\begin{figure}
 \includegraphics[height=5cm]{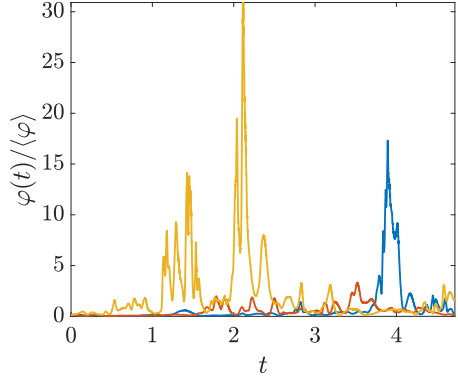}
\caption{Temporal evolution of the pseudo-dissipation $\varphi$ along 3 particle trajectories from DNS of \cite{Toschi2009}.}
\label{3_realization_epsilon}
\end{figure}

However, their is some  inconsistency in this theory \cite{Landau1963,Monin1975,Yeung1989}: $C_0$ was found not to be universal but Reynolds-dependent. Furthermore this scaling could not be extended to higher order moments of the velocity increments because the instantaneous dissipation intermittently reaches very high values and so the global average of $\varepsilon$ is not the relevant scale. This is illustrated in Fig.~\ref{3_realization_epsilon} where the pseudo-dissipation along fluid particle paths $\varphi$, an analogous variable to $\varepsilon$ that we define later in Eq.~\ref{eq:varphi}, is plotted and exhibits brief and sudden high fluctuations. The long-range correlation of the dissipation indicates that the large scales of the flow influence the local dissipation rate thus raising the question of the universality of the flow \cite{Bos2019}. These remarks (raised in \cite{Landau1944}) led Kolmogorov and Obukhov to the refined similarity hypothesis with the consideration of a locally-averaged dissipation \cite{Kolmogorov1962}. The subscript $\tau$ represents the time-scale of the locally-averaged variable.
  \begin{equation}
\varepsilon_\tau(t) = \frac{1}{\tau} \int_t^{t+\tau} \varepsilon(s) \mathrm{d}s
\end{equation}
The refined similarity hypothesis of K62 state that the statistics of velocity increments $\Delta_\tau u$ conditioned by local dissipation $\varepsilon_\tau$ is universal:
\begin{equation}
\mean{[\Delta_\tau u]^p| \varepsilon_\tau} = C_p \tau^{p/2} \varepsilon_\tau^{p/2}
\end{equation}
The unconditional statistics of the velocity increments therefore depend on the statistics of the locally-averaged dissipation: 
\begin{equation}
\mean{[\Delta_\tau u]^p} = C_p \tau^{p/2} \mean{\varepsilon_\tau^{p/2}}
\end{equation}

Such velocity structure functions have been studied and characterized in \cite{Mordant2004,Xu2006,Biferale2008,Arneodo2008}.
In K62, it was also suggested a log-normal distribution for $\varepsilon_\tau$, with a logarithm scaling for the variance of $\log \varepsilon_\tau$:
\begin{equation}
\sigma^2_{\log \varepsilon_\tau} \sim \log \dfrac{T_L}{\tau}
\end{equation}
This prediction is in reasonable agreement with experimental data \cite{Mordant2002} and was also obtained in \cite{Yaglom1966} with the discrete cascade model. 
These equations are not technically formulated in Kolmogorov's theory, which is instead expressed in a Eulerian framework by spatially averaging $\varepsilon$. However, considering a time average along the trajectory of particles is a natural extension of K62 theory for Lagrangian increments and this formalism have already been adopted before \cite{Sawford2015,Chevillard2012,Borgas1993}. \\ 

\begin{figure}
 \includegraphics[height=5cm]{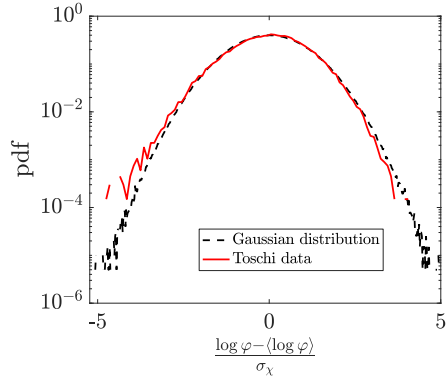}
\caption{PDF of normalized $\log \varphi$ compared to Gaussian distribution at different times.}
\label{PDF_chi_St0}
\end{figure}

The PDF of the pseudo-dissipation is in better agreement with the log-normal distribution than the classical dissipation, as shown in \cite{Pope1990}. It is defined as: 
\begin{equation}
\varphi(\mathbf{x},t) \equiv \nu \frac{\partial u_i}{\partial x_j} \frac{\partial u_i}{\partial x_j} 
\label{eq:varphi}
\end{equation}
For an isotropic flow, we have $\mean{\varphi} = \mean{\varepsilon}$. The Lagrangian variable is related to the Eulerian field by: $\varphi(t) = \varphi(\mathbf{x}_f(t),t)$, where $\mathbf{x}_f(t)$ denotes the position of a fluid particle at time $t$.
Figure~\ref{PDF_chi_St0} compares the pdf of $\log \varphi$ obtained from the data of \cite{Toschi2009} with a normal distribution and we find good agreement. As suggested in K62 and measured in DNS by \cite{Yeung2006b}, we have $\sigma^2_{\log \varphi} = A + B \log \mathrm{Re}$. 

The locally averaged dissipation (also called coarse-grained dissipation) can be defined by:
\begin{equation}
\varphi_\tau(t) =\frac{1}{\tau} \int_t^{t+\tau} \varphi(s) \diff s
\end{equation}
This is the Lagrangian equivalent for the dissipation averaged over a ball of size $\ell$ considered by \cite{Kolmogorov1962} in their refined similarity hypothesis. It is introduced in \cite{Borgas1993} to characterize multifractal scaling properties for flow with large Reynolds numbers. 
 
Numerous studies on data analysis of intermittency in turbulence reveal the multifractal nature of the pseudo-dissipaton. The seminal work of Frisch \cite{Frisch1995} to characterize intermittency based on Kolmogorov theories was followed among others by \cite{Chevillard2009,Chevillard2011, Chevillard2017a, Schmitt2001,Schmitt2003,Pereira2018}. Combining all the properties of intermittency mentioned  in their work, we suggest the following list of criteria for the pseudo-dissipation to exhibit intermittency.
\begin{tcolorbox}
\begin{enumerate}[label=(\roman*)]
\item  \label{list:K41} Kolmogorov 1941 scaling: $\mean{\varphi} = \nu \tau_\eta^{-2}$
\item \label{list:lognormal } Kolmogorov 1962: $\varphi$ is log-normal with $\sigma^2_{\log \varphi} \sim \log \dfrac{T_L}{\tau_\eta}$
\item\label{list:K62_onepoint_scaling}   Multiscaling of the one-point statistics: $\mean{\varphi^p} \sim \left( \dfrac{T_L}{\tau_\eta}\right)^{\xi(p)}$, where $\xi(p)$ is a non-linear function. 
\item \label{list:K62_twopoints_scaling} Power-law scaling for the coarse-grained dissipation, in the inertial range: for $\tau_\eta \ll \tau \ll T_L, \quad\mean{\varphi_\tau^p} \sim \left( \dfrac{T_L}{\tau} \right)^{\xi(p)}$
\end{enumerate}
\end{tcolorbox}
The last two points  \ref{list:K62_onepoint_scaling} and \ref{list:K62_twopoints_scaling} are precisely the main characteristics of multifractal systems, which were considered for the modeling of the dissipation.

\subsection{Modeling of the pseudo-dissipation}
\label{subsec:Modeling_dissipation}

\subsubsection{Multifractal models}
\label{subsubsec:Multifractal_processes}
In the Eulerian framework, discrete cascade models and later continuous random fields were developed. 
Yaglom proposed a model of multiplicative cascade where eddies can be seen as an ensemble of cells \cite{Yaglom1966}. The largest scale is represented by a unique cell of size $L$ and is then divided into smallest cells of size $\ell_1 = L/\lambda$ where $\lambda$ is the constant scale ratio of the cascade model. This process is repeated until the smallest scales are reached, with the subdivision $\ell_N = \eta = L/\lambda^N$. The energy is transferred from one cell generation to the next with a positive ratio given by a random variable $\alpha_{i}$ with $\mean{\alpha_i} = 1$ that are independent and identically distributed. We can define for each cell of size $\ell_n$ the energy dissipation rate through it: 
\begin{equation}
\varphi_{\ell_n} = \alpha_{1}\alpha_{2}...\alpha_{n} \mean{\varphi}
\label{eq:discrete_cascade}
\end{equation}

Following the independence of the random variables $\alpha_i$, it is straightforward to calculate the moments of any coarse-grained dissipation $\varphi_{\ell_n}$:
\begin{equation}
\begin{array}{ll}
\mean{(\varphi_{\ell_n} )^p} &= \displaystyle \mean{\varphi}^p \prod_{i=1}^n \mean{(\alpha_i)^p} \\
&= \mean{\varphi}^p \mean{\alpha_i^p}^n \\
&= \mean{\varphi}^p \left( \dfrac{L}{\ell_n}\right)^{\xi(p)}
\end{array}
\end{equation}
where we used $n = \log_\lambda (L/\ell_n)$ and $\xi(p) = \log_\lambda \mean{\alpha^p}$. 
Depending on the distribution of the $\alpha_i$, different forms of $\xi(p)$ are found (see Refs.~\cite{Frisch1978, Benzi1984, Meneveau1987}). Such construction ensures immediately \ref{list:K62_onepoint_scaling} and \ref{list:K62_twopoints_scaling}, and $\varphi = \varphi_{\eta} = \alpha_{1}\alpha_{2}...\alpha_{N} \mean{\varphi}$ is log-normal according to the Central Limit Theorem, assuming it applies.\\

Two main criticisms of these models are made in Ref.~\cite{Mandelbrot1968}. The first concerns the absence of spatio-temporal structure in these Eulerian representations of the dissipation fields which lacks causality, a necessary ingredient. Equivalent Lagrangian models were then proposed, following the formalism of Lagrangian intermittency developed in \cite{Borgas1993}. The model of \cite{Biferale1998} is defined via a multiplicative process of independent stationary random processes with given correlation times. Properties \ref{list:K62_onepoint_scaling} and \ref{list:K62_twopoints_scaling} can here again only be verified for a finite number of scales depending on the constant scale ratio of the model $\lambda$.\\

 This brings us to the second critic raised in Ref.~\cite{Mandelbrot1968} who suggested to consider continuous cascade models such as Gaussian Multiplicative Chaos \cite{Kahane1985, Robert2010} for which no arbitrary scale is chosen. Stochastic integrals can be interpreted as an infinite sum, with continuous values of scales. Taking the exponential of stochastic integrals  gives a “continuous product" instead of the discrete one defined in Eq.~\ref{eq:discrete_cascade}. Several models \cite{Schmitt2001, Muzy2002, Chevillard2017a, Pereira2018} are based on this formalism which allows to combine the continuous vision of a cascade and a causal structure of the process. Specific properties of the stochastic integrals must be defined to ensure intermittency of the dissipation and we present them in the following section.

\subsubsection{Gaussian Multiplicative Chaos}
\label{subsubsec:GMC}

This section therefore reviews the Gaussian Multiplicative Chaos (GMC) formalism, introduced by Kahane \cite{Kahane1985} which allows to build a process for the pseudo-dissipation $\varphi(t)$ and we show that it is in  in agreement with the criteria of intermittency defined in section~\ref{subsec:Characterization_Intermittency}. The GMC involves the following form for the pseudo-dissipation: 
\begin{equation}
\varphi (t) = \mean{\varphi} \exp(\chi_t)
\end{equation}
where $\chi_t$ is a Gaussian process of variance $\sigma_\chi^2$. Its mean ${\mu_\chi = -\frac{1}{2}\sigma_\chi^2}$ is determined with the constraint that $\mean{\exp (\chi_t) } = 1$. 
We can parameterize this process by a zero-average Gaussian process $X_t$ and the intermittency coefficient $\mu^\ell$:
\begin{equation}
\chi_t = \sqrt{\mu^\ell} X_t - \dfrac{\mu^\ell}{2} \sigma_X^2
\label{eq:normalization_chi}
\end{equation}
where $\mu^\ell$ is given by $\mu^\ell = \sigma_\chi^2/  \sigma_X^2$ and the process $X_t$ is constructed to be approximately log-correlated: 
\begin{equation}
\mean{X_t X_s} \approx	 \log_+ \dfrac{1}{|t-s|} + g(t,s)
\label{eq:log_autocorr}
\end{equation}
where $g$ is a bounded function and ${ \log_+(u) = \max(\log u,0)}$. The covariance kernel thus possesses a singularity and a standard approach consists in regularizing the distribution $X_t$ by applying a “cut-off", based on a small parameter $\tau_\eta$ such that, in the limit of $\tau_\eta \rightarrow 0$, $\varphi(t)$ is a GMC in a well-posed abstract framework. Further details and proof of convergence are derived in the complementary paper \cite{Goudenege2021} which rigorously formalizes the construction of such a process as a limit of $\tau_\eta$-regularized processes.\\

Let us show how the GMC is adapted to meet the intermittency criteria we have defined. \\
The mean value $\mean{\varphi}$ is chosen accordingly with requirement \ref{list:K41} and the formalism of this model (i.e. exponential of a Gaussian variable) naturally ensures the log-normality of $\varphi$. 
Based on this formalism, it is possible to derive the moments of the dissipation and the coarse-grained dissipation from the log-normal moments. Calculations are detailed in appendix~\ref{sec:Moments_dissipation_calculations}. We obtain  Eq.~\ref{eq:varphi_moments} for the moments of the dissipation:
\begin{equation*}
\mean{\varphi^p} = \mean{\varphi}^p \exp \left(\mu^\ell  p(p-1) \dfrac{\sigma_X^2}{2}  \right)
\end{equation*}
 We can show that prescribing $\sigma_X^2 \sim \log \dfrac{T_L}{\tau_\eta}$, which corresponds to the last part of requirement~\ref{list:lognormal }, readily ensures requirement~\ref{list:K62_onepoint_scaling}:
 \begin{equation}
 \begin{array}{ll}
 \mean{\varphi^p} &\sim \mean{\varphi}^p \exp \left( \frac{\mu^\ell}{2}p(p-1) \log \dfrac{T_L}{\tau_\eta} \right) \\
&\sim \left( \dfrac{T_L}{\tau_\eta} \right)^{\xi(p)}
 \end{array}
 \label{eq:moments_dissipation}
\end{equation}
with the non-linear scaling power law $\xi(p) = \frac{\mu^\ell}{2}p(p-1) $.\\
This scaling is consistent with the large-scale dependency (related to the Reynolds number) of the pseudo-dissipation. \\

Finally, the multifractal property of the coarse-grained dissipation \ref{list:K62_twopoints_scaling} is ensured by the log-correlated auto-correlation of $X_t$. Indeed, prescribing $\mean{X_t X_{t+\tau}} \sim \log \dfrac{T_L}{\tau}$ gives in Eq.~\ref{eq:varphi_tau_moments}:
\begin{equation}
\begin{array}{ll}
\mean{\varphi_\tau^p} &=  \mean{\varphi}^p  \displaystyle \int_{[0,1]^p}  \exp \left(  \mu^\ell \sum\limits_{i<j} \mean{X_{\tau s_i} X_{\tau s_j}} \right)  \prod\limits_{i=1}^p \diff s_i  \\
\dfrac{\mean{\varphi_\tau^p}  }{  \mean{\varphi}^p }&=  \displaystyle \int_{[0,1]^p} \exp \Big( \mu^\ell \sum\limits_{i<j} \log \frac{T_L}{\tau (s_j - s_i)} \Big)  \prod\limits_{k=1}^p \diff s_k  \\
&=  \displaystyle \int_{[0,1]^p}  \exp  \Big( \mu^\ell \frac{p(p-1)}{2}\log \frac{T_L}{\tau} - \mu^\ell\sum\limits_{i<j}^p \log  (s_j - s_i) \Big) \prod\limits_{k=1}^p \diff s_k \\
&= \left( \dfrac{T_L}{\tau} \right)^{\xi(p)}  \displaystyle \int_{[0,1]^p}   \prod\limits_{i<j}^p \dfrac{1}{(s_j - s_i)^{\mu^\ell}}  \prod\limits_{k=1}^p \diff s_k
\end{array}
\label{eq:moments_coarse-grained_dissipation}
\end{equation}
%\begin{equation}
%\begin{array}{ll}
%\dfrac{\mean{\varphi_\tau^p}  }{  \mean{\varphi}^p }&=  \displaystyle \int_{[0,1]^p} \hspace*{-0.6cm} \exp \Big( \mu^\ell \sum\limits_{i<j} \log \frac{T_L}{\tau (s_j - s_i)} \Big)  \prod\limits_{k=1}^p \diff s_k  \\
%&=  \displaystyle \int_{[0,1]^p} \hspace*{-0.6cm} \exp  \Big( \mu^\ell \frac{p(p-1)}{2}\log \frac{T_L}{\tau} \\
%& \displaystyle \hspace*{1.7cm}  - \mu^\ell\sum\limits_{i<j}^p \log  (s_j - s_i) \Big) \prod\limits_{k=1}^p \diff s_k \\
%&= \left( \dfrac{T_L}{\tau} \right)^{\xi(p)}  \displaystyle \int_{[0,1]^p}   \prod\limits_{i<j}^p \dfrac{1}{(s_j - s_i)^{\mu^\ell}}  \prod\limits_{k=1}^p \diff s_k
%\end{array}
%\label{eq:moments_coarse-grained_dissipation}
%\end{equation}
Taking the limit $\mathrm{Re}\rightarrow \infty$, the moments of $\varphi$ diverge in Eq.~\ref{eq:moments_dissipation} because $\varphi$ is correlated over the large energy containing scales, whereas at a given scale $\tau$, moments of $\varphi_\tau$ converge in Eq.~\ref{eq:moments_coarse-grained_dissipation}, fulfilling the statistical properties required by the K62 phenomenology. It becomes independent of the Reynolds number and behaves as power law at small scales. 

\begin{figure}
 \includegraphics[height=5cm]{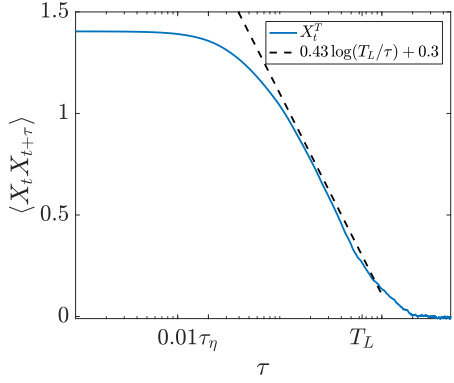}
\caption{Auto-correlation of $X_t^{T}$ full blue line compared with expected logarithmic behavior in black dotted line. }
\label{auto-correlation_X_Toschi}
\end{figure}

\subsubsection{Illustration from DNS data}
\label{subsubsec:Illustration}
We illustrate this logarithmic behavior of the auto-correlation on DNS realizations obtained by \cite{Toschi2009}. Their simulations were run at $\tau_\eta = 0.02$. The integral Lagrangian time is found to be $T_L = 0.64$ which gives, according to Ref.~\cite{Zhang2019a}, $\mathrm{Re}_\lambda = (T_L/\tau_\eta)/0.08 = 400$. For each fluid particle, the process $\chi_t$ is obtained by taking the logarithm of the pseudo-dissipation along the particle trajectory. $X_t^T$ is retrieved with the normalization of Eq.~\ref{eq:normalization_chi}, ignoring the multiplicative factor $\mu^\ell$ and we plot in Fig.~\ref{auto-correlation_X_Toschi} its auto-correlation. One can easily verified by comparison with the logarithmic behavior in dotted line that the auto-correlation follows a logarithmic behavior in the inertial range, i.e. between the $\tau_\eta$ and $T_L$ of the simulation. \\

\subsubsection{Conclusion}
A characterization of the intermittency has been proposed in section~\ref{subsec:Characterization_Intermittency} and we have checked that the proposed criteria are verified by a GMC modeling for the pseudo-dissipation. The remaining question to be addressed concerns the construction of the stochastic process $X_t$. Its variance must scale as the logarithm of the Reynolds number and its auto-correlation must be logarithmic in the inertial range. In the following, we summarize how such processes have been constructed in the literature. 

\subsection{Design of the  $X_t$ process}
\label{subsec:Long-range_correlation}
\begin{figure}
 \includegraphics[height=5cm]{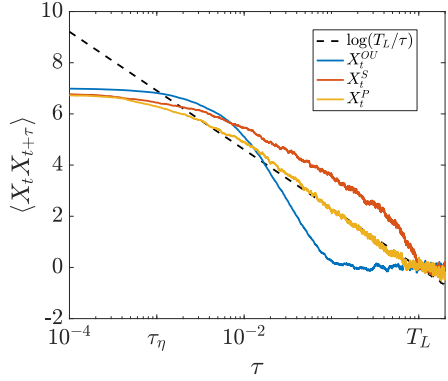}
\caption{Comparison of auto-correlation of processes of (a) Pope \cite{Pope1990} $X_t^{OU}$, (b) Schmitt \cite{Schmitt2003}  $X_t^S$  and (c) Pereira \cite{Pereira2018}  $X_t^P$ with logarithmic behavior. Processes are rescaled for comparable variance of $\log({T_L}/{\tau_\eta})$}
\label{auto-correlation_X_All}
\end{figure}

 Pope originally suggested to represent $\varphi$ as a log-normally correlated process by mean of Ornstein-Uhlenbeck process \cite{Pope1990}. 
The stochastic equation they proposed for $\chi(t) = \log \left( \varphi(t)/\langle \varphi \rangle \right) $ is the following: 
\begin{equation}
\diff \chi_t  = -\left( \chi_t + \frac{1}{2}\sigma_\chi^2 \right) \frac{\diff t}{T_\chi}+ \left( 2 \frac{\sigma_\chi^2 }{T_\chi} \right)^{1/2} \diff W_t
\label{eq:PopeOU}
\end{equation}
where $W_t$ is a Wiener process, $T_\chi$ is the integral time scale of $\chi$, extracted from DNS and found to be close to the Lagrangian Integral time scale $T_L$. The parameter $\sigma_\chi^2$ is Reynolds number dependent and is also chosen accordingly to DNS data.  The corresponding stochastic process $X_t^{OU}$ in this case is driven by:
\begin{equation}
\diff X_t^{OU} = -  X_t^{OU}  \frac{\diff t}{T_\chi} + \left( 2 \frac{\sigma_X^2 }{T_\chi} \right)^{1/2} \diff W_t
\end{equation}
The auto-correlation of this process is well-know and has an exponential decay in the form $\mean{X_t X_{t+\tau}} \sim \mathrm{e}^{-t/T_\chi}$. It is plotted in blue in Fig.~\ref{auto-correlation_X_All} and compared to the logarithmic behavior in the inertial range $[\tau_\eta, T_L] = [10^{-3}, 10^0]$. As expected, the exponential decay does not reproduce a long-range correlation. As a matter of fact, Pereira et al. showed in \cite{Pereira2018} that this model do not present the required multifractality for the coarse-grained process $\varphi_\tau(t)$. \\

Inspired by the stochastic process of Chevillard \cite{Chevillard2017a}, they proposed to replace the Ornstein-Uhlenbeck process by a fractional Ornstein-Uhlenbeck process, which  consists in replacing the Gaussian Noise $ \diff W_t$ in the Langevin equation by a fractional Gaussian noise $\diff W^H_t$.
Appropriate formalism for fractional Brownian motion (hereafter denoted fBm) was proposed in \cite{Mandelbrot1968}. They defined the fBm of exponent $H$ as a “moving average of $\diff W_t$, in which past increments of $W_t$, a Brownian motion, are weighted by the kernel $(t-s)^{H-1/2}$”. $H \in(0,1)$ is called the Hurst parameter and defines the roughness of the path. Standard Brownian motion corresponds to $H=1/2$ and is noted $W^{1/2}(t) = W_t$. 
A classic expression for the Holmgren-Riemann-Liouville fractional Brownian motion is the following one: 
\begin{equation}
W^H(t) = \dfrac{1}{\Gamma(H+1/2)} \displaystyle \int_0^t (t-s)^{H-1/2} \diff W_s
\label{eq:fBM}
\end{equation}

The particular case of Hurst parameter $H=0$ has a logarithmic auto-correlation \cite{Chevillard2017a} but as mentioned in section~\ref{subsubsec:GMC}, it is not well-defined because of the singularity of its auto-correlation in $0$. Mandelbrot \cite{Mandelbrot1968} proposed a regularization of this fBm: 
\begin{equation}
W^0_{\tau_\eta}(t) = \dfrac{1}{\sqrt{\pi}} \displaystyle \int_0^t (t-s+\tau_\eta)^{-1/2} \diff W_s
\label{eq:regularized_fBM}
\end{equation}
The calculation of its covariance gives, for any $t \geq s \geq 0$,
\begin{equation}
\begin{array}{ll}
\mean{W^0_{\tau}(t) W^0_{\tau}(s) } &= \displaystyle \int_{0}^{s } (t-u+\tau)^{-\frac{1}{2}} \, (s-u+\tau)^{-\frac{1}{2}}\, du \\
&= \displaystyle \int_{\tau}^{s+\tau} \frac{1}{\sqrt{u}\, \sqrt{t-s+u}}\, du \\
&= \displaystyle \left[ 2 \log \left(\sqrt{u} + \sqrt{u+t-s} \right) \right]_{\tau}^{s+\tau} \\
&=\displaystyle  2\log\left(\frac{\sqrt{s+\tau}+\sqrt{t+\tau}}{\sqrt{\tau}+\sqrt{\tau+|t-s|}}\right)
\end{array}
\end{equation}
It is shown in \cite{Goudenege2021} that the family of processes $\{ W^0_{\tau}(t) \}_{\tau>0}$ converges weakly in law to a Gaussian log-correlated process $W^0(t)$ with covariance expressed in Eq.~\ref{eq:log_autocorr}.\\

In his work, Schmitt \cite{Schmitt2001,Schmitt2003} developed the following causal stochastic process, inspired by this regularized process
\begin{equation}
X_t^S =  \int_{t+\tau_\eta - T_L}^t (t-s+\tau_\eta)^{-1/2}\diff W_s
\label{eq:Schmitt}
\end{equation}
 and showed its multifractal properties (scaling laws of the random process, of the coarse-grained process, logarithmic correlation of the logarithm of the procces etc.). 
 
Pereira  \cite{Pereira2018}, also used the increments of the regularized process in a fractional Ornstein-Uhlenbeck process to ensure stationarity of the process:
 \begin{equation}
\diff X_t^P=-\frac{1}{T_L} X_t^P \diff t+ \diff W^0_{\tau_\eta} (t)
\label{eq:Pereira}
\end{equation}

Figure~\ref{auto-correlation_X_All} compares the three processes described above with the logarithmic prediction of the auto-correlation in the inertial range ($[\tau_\eta , T_L] = [10^{-3}, 10^0]$). Both processes based on the fBm display long-range power-law correlation, as opposed to the Ornstein-Uhlenbeck process of Pope \cite{Pope1990}. \\

Subsequently, we will propose a new stochastic model for $X_t$, also inspired by a regularized fBm.  Beyond a purely mathematical construction of such a process, we would also like to introduce a natural physical interpretation before showing that it also allows a handy and efficient numerical implementation.

\section{Infinite sum of correlated Ornstein-Uhlenbeck processes}
\label{sec:Infinite_sum_OU}
We have seen that fBm present interesting auto-correlation properties with long-range behavior. The objective of this section is to propose a formalism different from that of Eq.~\ref{eq:fBM}, which does not involve a moving average because its simulation would require large memory. The expression we derive in the following, however, relies on a combination of Ornstein-Uhlenbeck processes which are markovian processes. This representation also makes it possible to make calculations (of moments and auto-correlation) easily, and to generate them by very simple calculation algorithms. 

\subsection{Approximation of fractional Brownian motion}
\label{subsec:Approximation_fBM}
\begin{figure}
 \includegraphics[height=6cm]{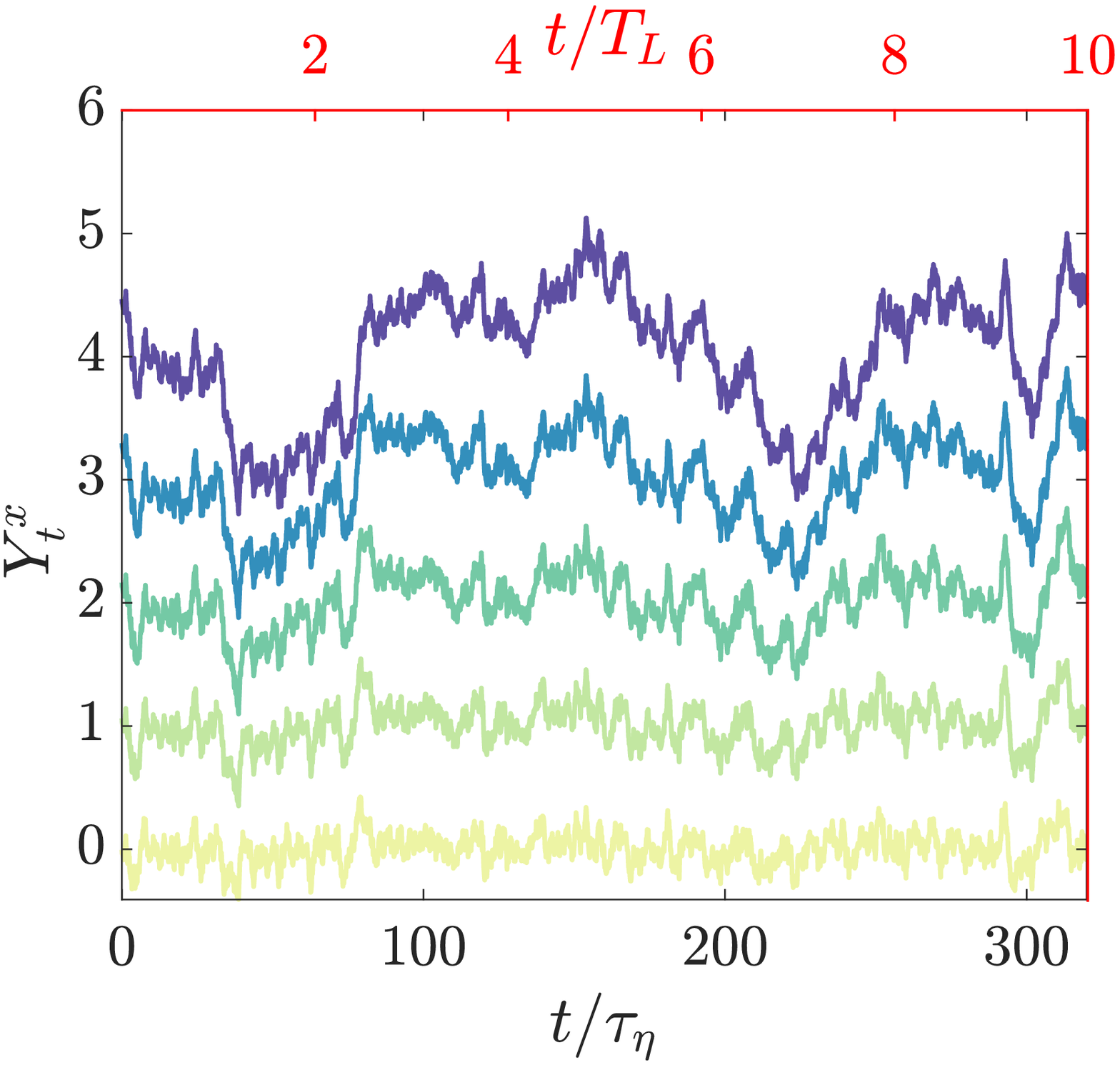}
\caption{5 correlated Ornstein-Uhlenbeck processes, driven by the same Wiener increments. Plots are shifted up for a better visualization and the color shades are darker with increasing characteristic times $x^{-1}$.}
\label{Ytx_vs_time_shifted}
\end{figure}

The fractional Brownian motion as defined in Eq.~\ref{eq:fBM} is a moving average and can be written in the following form:
\begin{equation}
B_t = \int_{0}^t K(t-s) \diff W_s
\end{equation}
Inspired by conventional techniques on Linear Time Invariant systems, we introduce the “spectral" representation of the kernel $K$. This transformation is also proposed in Ref~\cite{Carmona1998,Harms2020}. 
\begin{equation}
K(u) = \int_0^\infty e^{-ux} k(x) \diff x
\end{equation}
where $k$ is the inverse Laplace transform of $K$. 
If $K$ satisfies certain measurability properties, stochastic Fubini theorem allows us to exchange the two integrals after replacing the kernel by its spectral representation. 
\begin{equation}
\begin{array}{ll}
B_t	 	&=  \displaystyle\int_{0}^t K(t-s) \diff W_s \\
			&= \displaystyle\int_{0}^t  \left( \int_0^\infty e^{-x (t-s)} k(x) \diff x \right)  \diff W_s \\
			&= \displaystyle \int_0^\infty  \left(  \int_{0}^t  e^{-x (t-s)} \diff W_s \right)  k(x) \diff x \\
			&= \displaystyle \int_0^\infty \tilde{Y}_t^x  k(x) \diff x 
\end{array}
\label{eq:Infinite_sum_OU}
\end{equation}
where $\tilde{Y}_t^x = \int_{0}^t  e^{-x (t-s)} \diff W_s $ is a standardized Ornstein-Uhlenbeck process of parameter $x$ and initial value ${\tilde{Y}^x_0 = 0}$, solution of the stochastic differential equation: 
\begin{equation}
\diff \tilde{Y}_t^x = -x \tilde{Y}_t^x\diff t + \diff W_t
\label{eq:dYtx}
\end{equation}
Let us insist on the fact that all the Ornstein-Uhlenbeck processes appearing in the integrand are driven by the same Wiener increments $\diff W_t$ and are thus correlated to each other. Figure~\ref{Ytx_vs_time_shifted} shows $5$ correlated processes $Y_t^x$ with time-scales ranging from $\tau_\eta = 0.02$ to $T_L = 0.64$. The process $B_t$ is therefore a linear combination of standard Ornstein-Uhlenbeck processes, weighted by the kernel $k$.\\ 

We apply this technique to the fBm of Hurst parameter $H \in (0,1)$ as defined in Eq.~\ref{eq:fBM}. The corresponding kernel is $K(t) = \Gamma \left( H+ 1/2 \right)^{-1}  t^{H-1/2}$  and its inverse Laplace transform is $k(x) =  \Gamma \left( H+1/2 \right)^{-1} \Gamma \left( -H-1/2 \right)^{-1} x^{-H-1/2}$. The fBm can finally be written: 
\begin{equation}
W^H(t) = \dfrac{1}{\Gamma \left( H+1/2 \right) \Gamma \left( -H-1/2 \right) } \int_{0}^\infty \tilde{Y}_t^x x^{-H-1/2} \diff x 
\label{eq:fBM_OU}
\end{equation}
And introducing the increments of the Ornstein-Uhlenbeck processes defined in Eq.~\ref{eq:dYtx}, we readily obtain:
\begin{equation}
\diff W^H(t) \propto  \int_{0}^\infty \diff \tilde{Y}_t^x x^{-H-1/2} \diff x 
\label{eq:increments_fBM_OU}
\end{equation}

We have shown that the fBm can be expressed as an infinite sum of correlated Ornstein-Uhlenbeck processes, weighted by $k$, the inverse Laplace transform of the initial kernel function $K$ in the moving average of Eq.~\ref{eq:fBM}. This formulation has the advantage that no convolution product appears, and therefore the simulation of such a process does not require long-term memory. Inspired from this formalism, we propose a new process for $X_t$. 

\subsection{A new stochastic process with appropriate regularizations}
\label{subsec:Regularization}
As shown in section~\ref{subsec:Long-range_correlation}, fBm have been successfully used to reproduce multifractal properties and we therefore use the expression derived in Eq.~\ref{eq:Infinite_sum_OU} to suggest the following stochastic model for $X_t$:
\begin{equation}
X_t = \displaystyle \int_0^\infty Y^x_t k(x) \diff x
\label{eq:New_Xt}
\end{equation}
where $Y_t^x$ is an Ornstein-Uhlenbeck process of parameter $x$ and $k(x)$ has to be determined. 
We now give the constraints on such model to ensure the stationarity, the finite variance and the logarithmic auto-correlation of $X_t$.\\

\textbf{Stationarity}: A sufficient condition of stationarity for $X_t$ is to impose stationarity for all the Ornstein-Uhlenbeck processes $Y_t^x$. 
\begin{equation}
Y_t^x = \displaystyle \int_{-\infty}^t \mathrm{e}^{-x(t-s)} \diff W_s
\end{equation}

\textbf{Logarithmic auto-correlation}: 
The auto-correlation of this process is: 
\begin{equation}
\begin{array}{ll}
\mean{X_t X_{t+\tau}} &= \displaystyle \int_0^\infty \int_0^\infty  \mean{Y^x_t Y^y_{t+\tau}} k(x) k(y) \diff x \diff y \\
&= \displaystyle \int_0^\infty \int_0^\infty  \dfrac{\mathrm{e}^{-\tau y}}{x+y} k(x) k(y) \diff x \diff y 
\end{array}
\label{eq:auto-correlation_X}
\end{equation}
Where the term $\mean{Y^x_t Y^y_{t+\tau}}$ is developped in appendix~\ref{sec:N_points_correlation_OU}.  \\
We have seen that a fBm of Hurst $H=0$ has a logarithmic auto-correlation, at least approximately i.e., apart from the singularity. Based on the inverse Laplace transformation of the kernel $K(t) \sim t^{-1/2}$, we propose $k(x) \sim x^{-1/2}$. However, this kernel possesses a singularity at $0$ and we need to introduce regularizations to ensure a finite variance. \\

\textbf{Finite variance}: $X_t$ is zero-averaged and its auto-correlation function only depends on the delay $\tau$ because of stationarity. The variance of the process can be expressed as:
\begin{equation}
\displaystyle \int_0^\infty \int_0^\infty  \dfrac{k(x) k(y)}{x+y}  \diff x \diff y < \infty
\label{eq:finite_var}
\end{equation}

To satisfy and combine these three requirements, we propose to regularize the kernel $k$ in the following way. 
One can see on the auto-correlation of Eq.~\ref{eq:auto-correlation_X} that any contribution of the function $k(y)$ for $y \gg 1/\tau$ will vanish because of the term $\mathrm{e}^{-\tau y}$. Therefore, we introduce $\tau_\eta$ and we can assume $k(x) \sim x^{-1/2}$ only for $x  \ll \tau_\eta^{-1} $, which is now compliant with the integrability on $\mathbb{R}^+$. From a physical point of view, this regularization can be thought as a viscous cut-off.\\
A second regularization step is needed to ensure a finite variance of the process $X_t$, which corresponds to the need to introduce a large scale. More precisely, the second requirement \ref{list:K62_onepoint_scaling} specifies $\mean{X_t^2} \sim \log \frac{T_L}{\tau_\eta}$. It implies the integrability of $(x,y) \rightarrow \frac{k(x)k(y)}{x+y}$ on $(\mathbb{R^+})^2$ and the logarithmic behavior in the inertial range is ensured by the requirement of $k(x) \sim x^{-1/2}$ for $T_L^{-1} \ll x \ll \tau_\eta^{-1}$. \\ 

In light of these regularizations, we propose a new model for the process $X_t^\infty$. Note that we use the superscript “$\infty$" because it highlights the use of an infinite sum of Ornstein-Uhlenbeck processes. 
\begin{equation}
X_t^\infty = \displaystyle \int_0^\infty Y^x_t \dfrac{1}{\sqrt{x}} \left( g_{T_L}(x) - g_{\tau_\eta}(x) \right) \diff x
\label{eq:New_Xt_regularized}
\end{equation}
where $g$ is such that the integral defined by the auto-correlation in Eq.~\ref{eq:finite_var} converges. A sufficient condition would be: 
\begin{equation}
g_\alpha(x) \rightarrow 
\left\lbrace
\begin{array}{ll}
0 & \text{if } x \ll 1/\alpha \\
1 & \text{if } x \gg 1/\alpha 
\end{array}
\right.
\end{equation}
The calculations of the auto-correlation is given in appendix~\ref{sec:Calculations_Xt}  where we show that $\mean{(X_t^\infty)^2} \sim \log \dfrac{T_L}{\tau_\eta}$ and $\mean{X_t^\infty  X^\infty_{t+\tau}} \sim  \log \dfrac{T_L}{\tau}$. \\

\begin{figure}
 \includegraphics[height=5cm]{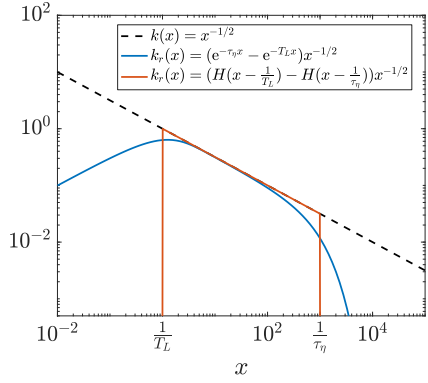}
\caption{Possible regularizations}
\label{Regularizations_hx}
\end{figure}

Examples of possible regularizations of the kernel $k(x) \sim x^{-1/2}$ are shown in Fig.~\ref{Regularizations_hx}: Cutting functions are $g_\alpha(x) = 1-\mathrm{e}^{-\alpha x}$ or $g_\alpha(x) = H(x-1/\alpha)$ where $H$ is the heaviside function. 

The following section presents other type of regularizations in the spectral representation that can lead to existing processes. 

\subsection{A framework encompassing existing processes}
\label{subsec:Framework_encompassing_processes}

In this section, we show that previous stochastic processes can be obtained from Eq.~\ref{eq:New_Xt} with appropriate regularizations. 
The regularized fBm introduced in Ref.~\cite{Mandelbrot1968} in Eq.~\ref{eq:regularized_fBM} can be expressed as an infinite sum of Ornstein-Uhlenbeck processes, introducing the inverse Laplace transorm and applying the Fubini's theorem (as it is done in section~\ref{subsec:Approximation_fBM}). The inverse Laplace transform of $K(t) =\pi^{-1/2} (t+\tau_\eta)$ is $k_{\tau_\eta}(x) =\pi^{-1} \mathrm{e}^{-\tau_\eta x} x^{-1/2}$. We obtain: 
\begin{equation}
W^0_{\tau_\eta}(t) = \dfrac{1}{\sqrt{\pi}} \displaystyle \int_{0}^t (t-s+\tau_\eta)^{-1/2} \diff W_s = \dfrac{1}{\pi} \displaystyle \int_{0}^\infty \tilde{Y}_t^x k_{\tau_\eta}(x)\diff x 
\end{equation}
This exponential cutting function $\mathrm{e}^{-\tau_\eta x}$ allows the process to be well-defined, as opposed to the fBm of Hurst $0$. \\
This process is now well-defined, and with logarithmic auto-correlation, inherited from the behavior of the kernel $\sim x^{-1/2}$.  However, it is not stationary and no large scales have been introduced: its variance is $\log \frac{t+\tau_\eta}{\tau_\eta}$. This is why Pereira \cite{Pereira2018} and Schmitt \cite{Schmitt2003} introduced another regularization on the process. \\

\begin{table*}
\begin{ruledtabular}
\begin{tabular}{ccc}
 Process & Definition &Spectral representation \\ \hline
fBm: $ W^0_{{\tau_\eta}}(t)$ &   $\displaystyle \dfrac{1}{\sqrt{\pi}}  \int_0^t (t-s+{\tau_\eta})^{-1/2} \diff W_s$ &   $\displaystyle \dfrac{1}{\pi} \int_0^\infty Y_t^x \dfrac{\textcolor{red}{\mathrm{e}^{-\tau_\eta x}}}{ \textcolor{olive}{\sqrt{x}}} \diff x$ \\
Schmitt: $X_t^S $ & $ \displaystyle \int_{t+\tau_\eta - T_L}^t (t-s+\tau_\eta)^{-1/2}\diff W_s$ & $\displaystyle  \int_0^\infty \textcolor{blue}{\left(  \int_{t+\tau_\eta \textcolor{red}{- T_L}}^t\mathrm{e}^{-(t-s)x} \diff W_s \right)} \dfrac{\textcolor{red}{\mathrm{e}^{-\tau_\eta x}}}{ \textcolor{olive}{\sqrt{x}}} \diff x$ \\
Pereira: $X_t^P$ & $\displaystyle \int_{-\infty}^t  \mathrm{e}^{-(t-s)/T_L} \diff W^0_{\tau_\eta} (t) $ & $\displaystyle \int_{0}^\infty \textcolor{blue}{\left( \int_{-\infty}^t \textcolor{red}{\mathrm{e}^{-(t-s)/T_L}} \diff Y_s^x \right)} \dfrac{\textcolor{red}{\mathrm{e}^{-\tau_\eta x}}}{ \textcolor{olive}{\sqrt{x}}}\diff x$ \\
Pope: $X_t^{OU} $ & $ \displaystyle  \int_{-\infty}^t \omega  \mathrm{e}^{-(t-s)/T_\chi} \diff W_s $ & $\displaystyle \int_0^\infty  \textcolor{blue}{Y_t^{x}} \omega   \textcolor{red}{\delta(x-T_\chi^{-1})} \diff x$ \\
$ X_t^{\infty} $ & $\displaystyle \int_{-\infty}^t \left( t-s+\tau_\eta \right)^{-1/2} - \left( t-s+T_L \right)^{-1/2} \diff W_s$  & $ \displaystyle \int_0^\infty \textcolor{blue}{Y_t^{x}}  \dfrac{\left(   \textcolor{red}{g_{T_L}(x)} -   \textcolor{red}{g_{\tau_\eta}(x)} \right)}{\textcolor{olive}{\sqrt{x}}}  \diff x $ \\
\end{tabular}
\end{ruledtabular}
\caption{\label{table:summarize_regularization}Regularizations applied on the spectral representation of different processes. The logarithmic behavior of the auto-correlation of the process comes from the kernel behavior $x^{-1/2}$ in brown, its stationarity comes from the blue term and the red terms ensures the finite variance. }
\end{table*}

Let us examine the process defined by Schmitt \cite{Schmitt2003} and apply once again the same procedure than in section~\ref{subsec:Approximation_fBM}. For clarity, we omit the scaling factor $\pi^{-1/2}$.
\begin{equation}
\begin{array}{ll}
X_t^S &= \displaystyle \int_{t+\tau_\eta - T_L}^t (t-s+\tau_\eta)^{-1/2}\diff W_s \\
&= \displaystyle \int_{t+\tau_\eta - T_L}^t \left( \int_0^\infty \mathrm{e}^{-(t-s)x} k_{\tau_\eta}(x) \diff x \right) \diff W_s \\
&= \displaystyle  \int_0^\infty \left(  \int_{t+\tau_\eta - T_L}^t\mathrm{e}^{-(t-s)x} \diff W_s \right) k_{\tau_\eta}(x) \diff x 
\end{array}
\end{equation}
To ensure finiteness of the variance, the Ornstein-Uhlenbeck processes must have a finite memory. In this case, the integral is truncated and the regularization consists in replacing the Ornstein-Uhlenbeck $Y_t^x$ of Eq.~\ref{eq:New_Xt}, by $\int_{t-T_L+\tau_\eta}^t e^{-x (t-s)} \diff W_s$. \\ 

The process of Pereira \cite{Pereira2018} is based on the increments of the regularized fBm: 
First, it is interesting to show that we can easily retrieve Chevillard's expressio of increments when applying the technique to the infinitesimal increment of $W^0_{\tau_\eta}$  \cite{Chevillard2017a}:
\begin{equation}
\begin{array}{ll}
\diff W^0_{\tau_\eta} &= \displaystyle \int_{0}^\infty \diff \tilde{Y}_t^x k_{\tau_\eta}(x)\diff x \\
&= \displaystyle \int_{0}^\infty (-x \tilde{Y}_t^x \diff t + \diff W_t) k_{\tau_\eta}(x)\diff x \\
&= \displaystyle \int_{0}^\infty -x k_{\tau_\eta}(x) \int_{0}^t \mathrm{e}^{-(t-s)x} \diff W_s \diff x \diff t + \displaystyle \int_{0}^\infty k_{\tau_\eta}(x) \diff x \diff W_t \\
\end{array}
\end{equation}
%\begin{equation}
%\begin{array}{ll}
%\diff W^0_{\tau_\eta} &= \displaystyle \int_{0}^\infty \diff \tilde{Y}_t^x k_{\tau_\eta}(x)\diff x \\
%&= \displaystyle \int_{0}^\infty (-x \tilde{Y}_t^x \diff t + \diff W_t) k_{\tau_\eta}(x)\diff x \\
%&= \displaystyle \int_{0}^\infty -x k_{\tau_\eta}(x) \int_{0}^t \mathrm{e}^{-(t-s)x} \diff W_s \diff x \diff t \\
%&+ \displaystyle \int_{0}^\infty k_{\tau_\eta}(x) \diff x \diff W_t \\
%\end{array}
%\end{equation}
Using the  Laplace transform for the first integral, we have: $\mathcal{L} \left( x k_{\tau_\eta}(x) \right) = \mathcal{L} \left( \sqrt{x} \mathrm{e}^{-\tau_\eta x} \right) = \Gamma \left( {3}/{2}\right) t^{-3/2}$. And the second integral is $\int_{0}^\infty k_{\tau_\eta}(x) \diff x = \tau_\eta^{-1/2}$. This yields:
\begin{equation}
\begin{array}{ll}
\diff W^0_{\tau_\eta} &= \displaystyle \int_{0}^t \dfrac{-1}{2} (t-s+\tau_\eta)^{-3/2} \diff W_s \diff t +  \tau_\eta^{-1/2} \diff W_t \\
&= \tilde{\beta}_{\tau_\eta}(t) \diff t +  \tau_\eta^{-1/2}\diff W_t
\end{array}
\end{equation}
where $ \tilde{\beta}_{\tau_\eta}(t) = \dfrac{-1}{2} \displaystyle \int_{0}^t (t-s+\tau_\eta)^{-3/2} \diff W_s $. 
In \cite{Pereira2018} they rather use the stationary version of this increment with $\beta_{\tau_\eta}(t) = \dfrac{-1}{2} \displaystyle \int_{-\infty}^t (t-s+\tau_\eta)^{-3/2} \diff W_s$ and based on that increment, they regularize the process with:
\begin{equation}
\begin{array}{ll}
X_t^P &= \displaystyle \int_{-\infty}^t \mathrm{e}^{-(t-s)/T_L} \diff W^0_{\tau_\eta}(s) \\
&= \displaystyle \int_{-\infty}^t \mathrm{e}^{-(t-s)/T_L}  \int_{0}^\infty \diff Y_s^x k_{\tau_\eta}(x) \diff x  \\
&= \displaystyle \int_{0}^\infty \left( \int_{-\infty}^t \mathrm{e}^{-(t-s)/T_L} \diff Y_s^x \right) k_{\tau_\eta}(x) \diff x 
\end{array}
\end{equation}
Therefore, the regularization for stationarity consists in replacing the Ornstein-Uhlenbeck $Y_t^x$ of Eq.~\ref{eq:New_Xt}, by $\int_{-\infty}^t \mathrm{e}^{-(t-s)/T_L} \diff Y_s^x$. \\

Finally, we can see that taking a Dirac function for $k(x)$ corresponds to Pope's process \cite{Pope1990}: 
\begin{equation}
X_t^{OU}  = \displaystyle \int_{-\infty}^t \omega  \mathrm{e}^{-(t-s)/T_\chi} \diff W_s  =  \int_0^\infty Y_t^{x} \omega   \delta(x-T_\chi^{-1}) \diff x
\end{equation}
where $\omega = \left( 2 \frac{\sigma_\chi^2 }{T_\chi} \right)^{1/2} $ is the scaling factor in front of the Gaussian Noise in Eq.~\ref{eq:PopeOU}. 
This kernel representation does not exhibit a behavior in $x^{-1/2}$, and we already know that the single Ornstein-Uhlenbeck process is not log-correlated. \\

Table~\ref{table:summarize_regularization} summarizes the different regularizations for all these processes. It shows that the general formalism of Eq.~\ref{eq:New_Xt} is a framework that encompasses existing processes depending on the three criteria for the regularization. 

If the general formalism proposed in Eq.~\ref{eq:New_Xt} gives the possibility to represent and simulate these processes using Ornstein-Uhlenbeck processes, one can see that their simulation is not equivalent. Processes of Schmitt \cite{Schmitt2003} and Pereira  \cite{Pereira2018} require to keep in memory the history of the process since at each instant $t$, the set of realizations of $W_s$ and $Y_s^x$ respectively, for $s$ in the intervals $[t+\tau_\eta-T_L, t]$ and $]-\infty,t]$ respectively should be involved in the computation (it is actually truncated for the numerical simulation). It is not the case for the one we propose in Eq.~\ref{eq:New_Xt_regularized} and we develop in the following section a numerical approach to implement such process with no long-term memory.

\FloatBarrier

\section{Finite sum of correlated Ornstein-Uhlenbeck processes}
\label{sec:Finite_sum_OU}

\subsection{Quadrature}
\label{subsec:quadrature}
\begin{figure}
 \includegraphics[height=5cm]{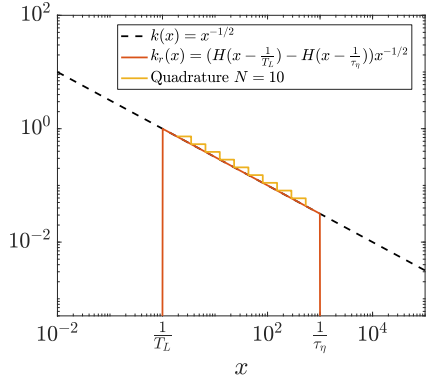}
 \caption{\label{Quadrature_hx} The kernel behavior $k(x) \sim x^{-1/2}$ in dotted line is regularized with Heaviside cutting functions in red and compared to its quadrature representation in yellow.}
\end{figure}

Following the idea of \cite{Harms2020}, with an appropriate quadrature, the integral can be replaced by a system of finite number of Ornstein-Uhlenbeck processes. We call $X_t^\infty$ the process defined with the infinite sum and $X_t^N$ the one obtained with $N$ points of quadrature.

\begin{equation}
X_t ^\infty \equiv \displaystyle \int_0^\infty Y^x_t \dfrac{1}{\sqrt{x}} \left( g_{T_L}(x) - g_{\tau_\eta}(x) \right) \diff x \approx X_t^N \equiv  \sum_{i=1}^N \omega_i Y^{x_i}_t
\end{equation}
Because of the regularizing functions $g_{T_L} - g_{\tau_\eta}$ , it is useless to compute quadrature points far outside the inertial range $[T_L^{-1};\tau_\eta^{-1}]$.
For simplicity, we use in the following examples Heaviside functions for $g$. 
 Considering the logarithmic shape of the kernel, we propose a geometric partition of this domain, along with a middle-Riemann sum for the weights: 
\begin{equation}
\text{for} \quad  i=1,...,N_{x_i}  
\left\lbrace
\begin{array}{ll}
x_i &= \dfrac{1}{T_L} \left( \dfrac{T_L}{\tau_\eta}\right)^{\frac{i-1/2}{N_{x_i}}} \\
\omega_i &= \dfrac{1}{\sqrt{x_i}} \Delta x_i 
\end{array}
\right.
\label{eq:Quadrature}
\end{equation}
Where $\Delta x_i =  \dfrac{1}{T_L} \left( \dfrac{T_L}{\tau_\eta}\right)^{\frac{i}{N_{x_i}}} -  \dfrac{1}{T_L} \left( \dfrac{T_L}{\tau_\eta}\right)^{\frac{i-1}{N_{x_i}}} $. 
Figure~\ref{Quadrature_hx} shows the kernel approximation with $N=10$ points of quadrature. The kernel $x^{-1/2}$ is approached by step functions all along the inertial range. The weights can be normalized to match the variance of the analytic process $\mean{(X_t^\infty)^2}$. The normalizing factor $R$ is given by:
\begin{equation}
R = \dfrac{\sigma_{X_t^\infty}}{\sigma_{X_t^N}} = \sqrt{\mean{(X_t^\infty)^2}} \left( \displaystyle  \sum_{i=1}^N \dfrac{\omega_i \omega_j}{x_i + x_j} \right)^{-1/2}
\end{equation}

\begin{figure}
  \includegraphics[height=5cm]{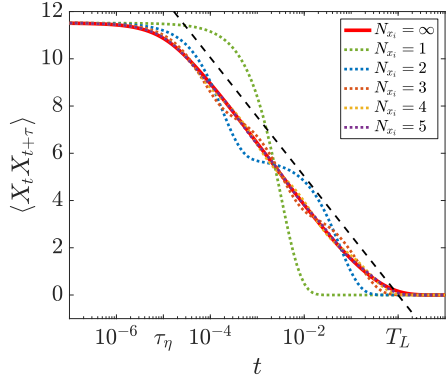}
\caption{\label{Comparison_Nxi_5decades} Comparison of the auto-correlations of the analytical process and the discrete one for a finite number of modes. The inertial range covers $5$ decades. All the processes are normalized by a unit variance. }
\end{figure}

Figure~\ref{Comparison_Nxi_5decades} shows the auto-correlation of the process $X_t^\infty$ compared with the discrete one $X_t^N$. As demonstrated in appendix~\ref{sec:Calculations_Xt}, it is clear that the infinite sum has indeed a logarithmic auto-correlation, it follows the dotted line all along the inertial range. A one point quadrature, corresponding to a single Ornstein-Uhlenbeck process is plotted in green in the figure. As discussed above, this specific process corresponds to $X_t^1 = X_t^{OU}$ and does not have a long-range correlation all along the inertial range. With two points of quadrature, the auto-correlation, in blue, displays two bumps, around the two time-scales of the Ornstein-Uhlenbeck processes. The auto-correlation range has been extended but it is not yet clear that it follows a logarithmic behavior. With more quadrature points, the auto-correlation of $X_t^N$ is getting closer to the analytical one. \\

\begin{figure}
 \includegraphics[height=5cm]{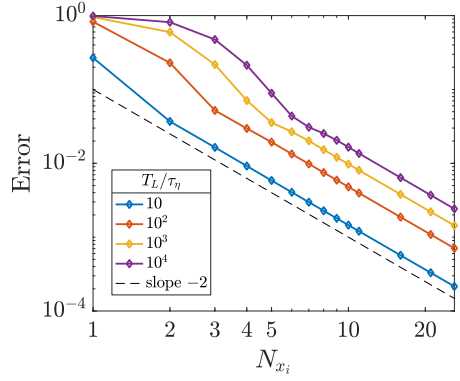}
 \caption{\label{Relative_error_quadrature} Numerical convergence of the auto-correlation of $X_t^{N_{x_i}}$ towards the expected correlation of $X_t^\infty$.}
\end{figure}

\begin{figure}
  \includegraphics[height=5cm]{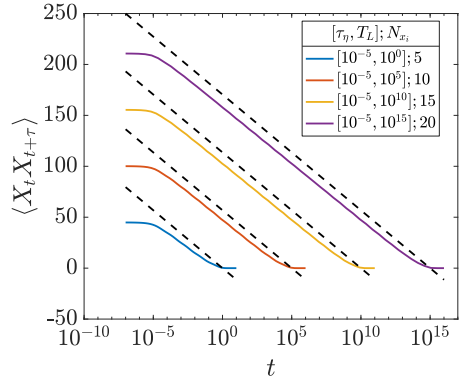}
\caption{\label{Quadrature_1OU_per_decade} Auto-correlation of $X_t$ for different inertial ranges. The number of points chosen in the quadrature corresponds to the number of decades covered by the inertial range.}
\end{figure}

This convergence of the auto-correlation can be explicit introducing the relative difference between the analytical auto-correlation $\rho^\infty(\tau)$ and the one obtained from the quadrature $\rho^N(\tau)$: 
\begin{equation}
\begin{array}{ll}
\rho^\infty(\tau) &\equiv \displaystyle \int_{T_L^{-1}}^{\tau_\eta^{-1}} \int_{T_L^{-1}}^{\tau_\eta^{-1}}   f(\tau,x,y) \diff x \diff y \\
 \rho^N(\tau) &\equiv \displaystyle  \sum_{i=1}^N \sum_{j=1}^N f(\tau,x_i,x_j) \Delta x_i \Delta x_j
\end{array}
\end{equation}
where $f(\tau,x,y) = \dfrac{\mathrm{e}^{-\tau y}}{(x+y) \sqrt{xy}}$. The numerical convergence is verified in Fig.~\ref{Relative_error_quadrature} with the error defined as: 
\begin{equation}
\mathrm{Error} = \displaystyle \sqrt{\int_{\tau_\eta}^{T_L} \left( \dfrac{\rho^\infty(\tau) - \rho^N(\tau)}{\rho^\infty(\tau)}\right)^2 \diff \tau }
\end{equation}
The order of convergence  is $2$. The value of the error is shifted when increasing the inertial range. With one Ornstein-Uhlenbeck per decade, the relative error is below $10\%$.  We can therefore postulate that an acceptable number of processes would be one or two per decade. Figure \ref{Quadrature_1OU_per_decade} illustrates this choice, with different inertial ranges. The number of points for the discrete process is chosen accordingly and we verify the logarithmic behavior of such processes all along the inertial range. For instance, with an inertial range covering 20 decades (purple line in Fig.~\ref{Quadrature_1OU_per_decade}), that corresponds to $\mathrm{Re}_\lambda \sim 10^{11}$, only 20 Ornstein-Uhlenbeck processes are needed to approach a logarithmic behavior of the auto-correlation all along the inertial range. 

\subsection{Discussion}
\label{subsec:Discussion}

A new log-correlated process $X_t^\infty$ and its discrete version with $N$ Ornstein-Uhlenbeck processes $X_t^N$ have been presented. 
In this section, we will discuss their physical interpretation and their advantage over existing processes. 

\subsubsection{Physical interpretation}
$X_t^N$ can be seen as an extension of Pope's process \cite{Pope1990}. We recall that the latter corresponds to $N=1$ with quadrature points taken as 
\begin{equation}
x_1 = \dfrac{1}{T_\chi}, \quad \text{ and }
\omega_1 = \sqrt{\dfrac{2 \sigma_\chi^2}{\mu^\ell T_\chi}} 
\end{equation}
\cite{Pope1990} observed that $T_\chi$ scales with the integral time scale $T_L$ and $\sigma_\chi$ scales with logarithm of Reynolds number. By comparison with our proposition of quadrature, we would suggest to use: $T_\chi = \sqrt{T_L \tau_\eta}$, and $\sigma_\chi$ is indeed scaling as $\sigma_\chi \sim \log \frac{T_L}{\tau_\eta}$ to ensure requirement \ref{list:K62_onepoint_scaling}. \\
For Reynolds number $\mathrm{Re}_\lambda \sim \frac{T_L}{0.08 \tau_\eta} \lessapprox 125$, we have seen that a single Ornstein-Uhlenbeck is enough to cover the entire inertial range and the exponential decay mimics the logarithmic behavior in such small interval. However, for larger Reynolds number,  it is necessary to extend the long-range of the auto-correlation by adding other Ornstein-Uhlenbeck processes, evenly distributed all along the inertial range. A perfect logarithmic scaling is retrieved with an infinity of Ornstein-Uhlenbeck processes. \\

This new process also makes a very simple link between “continuous" processes with no time scale (or here, an infinity), corresponding to $X_t^\infty $ and “discrete” cascade models $X_t^N$, where arbitrary time-scales are chosen to each represent a turbulent structure. A turbulent cascade is often represented as a product of independent processes defined at each scale, each one presenting a characteristic time scale. The approximation of $X_t^\infty$ by $X_t^N$ exactly consists in selecting representative time-scales, and the coherence of the whole cascade is ensured by the fact that every Ornstein-Uhlenbeck process is correlated to each other because driven by the exact same Gaussian Noise. 

\subsubsection{Implementation}
 Unlike the models of \cite{Schmitt2003, Pereira2018}, the process has no 'self-memory'. It is the combination of several Ornstein-Uhlenbeck processes, with adapted characteristic time-scales that can mimic this long-range correlation. The closer are the characteristic times of the Ornstein-Uhlenbeck processes, the better is the logarithmic approximation (quadrature with a large number of points) but we show that one time scale per decade is already enough to retrieve the approximate long-range behavior. This considerably reduce the computational cost of the simulation of such process. Simulating Ornstein-Uhlenbeck processes is very common, rapid and does not require to keep a memory of the history of the path, as opposed to the convolution form used in Refs.~\cite{Schmitt2003, Pereira2018}. 

%\begin{widetext}
\subsubsection{A causal multifractal process for pseudo-dissipation}
An analytical stochastic equation can be derived for the pseudo-dissipation, which is the variable of interest used in Lagrangian stochastic models. First, we can retrieve an analogous formulation for the incremenents of $X_t^\infty$, introducing the $\beta$ function already used in \cite{Pereira2018}. 

%\begin{equation}
%\begin{array}{ll}
%d X_t^\infty &= \displaystyle \int_0^\infty dY_t^x \dfrac{\mathrm{e^{-x\tau_\eta}-e^{-x T_L}}}{\sqrt{x}} \diff x \\
%&=  \displaystyle\int_0^\infty (-x Y_t^x \diff t + \diff W_t) \dfrac{\mathrm{e^{-x\tau_\eta}-e^{-x T_L}}}{\sqrt{x}} \diff x \\
%&=  \displaystyle \dfrac{-1}{2} \int_{-\infty}^t \Big( (t-s+\tau_\eta)^{-3/2} \\
%&  \quad \quad \quad \quad \quad - (t-s+T_L)^{-3/2} \Big) \diff W_s \diff t \\
%& \quad \quad \quad  + \left(\dfrac{1}{\sqrt{\tau_\eta}} - \dfrac{1}{\sqrt{T_L}}\right) \diff W_t \\
%&= \left( \beta_{\tau_\eta}(t) - \beta_{T_L}(t)\right) \diff t + \left(\dfrac{1}{\sqrt{\tau_\eta}} - \dfrac{1}{\sqrt{T_L}}\right) \diff W_t 
%\end{array}
%\end{equation}

\begin{equation}
\begin{array}{ll}
\diff X_t^\infty &= \displaystyle \int_0^\infty \diff Y_t^x \dfrac{\mathrm{e^{-x\tau_\eta}-e^{-x T_L}}}{\sqrt{x}} \diff x \\
&=  \displaystyle\int_0^\infty (-x Y_t^x \diff t + \diff W_t) \dfrac{\mathrm{e^{-x\tau_\eta}-e^{-x T_L}}}{\sqrt{x}} \diff x \\
&=  \displaystyle \dfrac{-1}{2} \int_{-\infty}^t \Big( (t-s+\tau_\eta)^{-3/2} - (t-s+T_L)^{-3/2} \Big) \diff W_s \diff t \left(\dfrac{1}{\sqrt{\tau_\eta}} - \dfrac{1}{\sqrt{T_L}}\right) \diff W_t \\
&= \left( \beta_{\tau_\eta}(t) - \beta_{T_L}(t)\right) \diff t + \left(\dfrac{1}{\sqrt{\tau_\eta}} - \dfrac{1}{\sqrt{T_L}}\right) \diff W_t 
\end{array}
\end{equation}
%\begin{equation}
%\begin{array}{ll}
%d X_t^\infty &= \displaystyle \int_0^\infty dY_t^x \dfrac{\mathrm{e^{-x\tau_\eta}-e^{-x T_L}}}{\sqrt{x}} \diff x \\
%&=  \displaystyle\int_0^\infty (-x Y_t^x \diff t + \diff W_t) \dfrac{\mathrm{e^{-x\tau_\eta}-e^{-x T_L}}}{\sqrt{x}} \diff x \\
%&=  \displaystyle \dfrac{-1}{2} \int_{-\infty}^t \Big( (t-s+\tau_\eta)^{-3/2} \\
%& \hspace*{3cm} - (t-s+T_L)^{-3/2} \Big) \diff W_s \diff t \\
%& \hspace*{1cm} + \left(\dfrac{1}{\sqrt{\tau_\eta}} - \dfrac{1}{\sqrt{T_L}}\right) \diff W_t \\
%&= \left( \beta_{\tau_\eta}(t) - \beta_{T_L}(t)\right) \diff t + \left(\dfrac{1}{\sqrt{\tau_\eta}} - \dfrac{1}{\sqrt{T_L}}\right) \diff W_t 
%\end{array}
%\end{equation}

We recall that the Lagrangian multiplicative chaos, which is causal and stationary, is readily obtained while exponentiating the Gaussian process $X_t^\infty$: ${\varphi = \mean{\varphi} \exp \left( \sqrt{\mu^\ell} X_t^\infty - \dfrac{\mu^\ell \sigma_X^2}{2}\right)}$.
Application of Ito’s lemma gives the Lagrangian stochastic dynamics of the pseudo-dissipation, namely:
\begin{equation}
\begin{array}{l}
\dfrac{\diff \varphi }{\varphi} = \Big[ \sqrt{\mu^\ell} \left(\beta_{\tau_\eta}(t) - \beta_{T_L}(t)\right) + \dfrac{\mu^\ell}{2}\left(\dfrac{1}{\sqrt{\tau_\eta}} - \dfrac{1}{\sqrt{T_L}}\right)^2 \Big] \diff t \\
\hspace*{1cm} + \sqrt{\mu^\ell} \left(\dfrac{1}{\sqrt{\tau_\eta}} - \dfrac{1}{\sqrt{T_L}}\right) \diff W_t
\end{array}
\label{eq:sto_pseudo-dissipation}
\end{equation}

Of course, the implementation of this stochastic equation preferentially uses the expression of $\beta$ in the Laplace domain and we recall that the same Wiener process is used in the $N$ Ornstein-Uhlenbeck processes $Y_t^{x_i}$ but also in $\diff W_t$ in Eq.~\ref{eq:sto_pseudo-dissipation}. Numerically, we replace the $\beta$-functions by its quadrature: 
\begin{equation}
\begin{array}{ll}
\beta_{\tau_\eta}(t) - \beta_{T_L}(t) &= \displaystyle \int_0^\infty -x Y_t^x \dfrac{g_{T_L}(x) - g_{\tau_\eta}(x)}{\sqrt{x}}  \diff x \\ 
& \displaystyle \approx \sum_{i=1}^N -x_i \omega_i Y_t^{x_i} 
\end{array}
\end{equation}

%\end{widetext}

\section{Conclusion}
\label{sec:Conclusion}

Intermittency in turbulence can be characterized by multifractal properties of the dissipation. A Gaussian Multiplicative Chaos formalism allows us to model such dissipation process, but relies on the introduction of a zero-average Gaussian and log-correlated process, $X_t$. In the literature, such processes were defined based on a regularized fBm ; they lack physical interpretation and can be computationally expensive in simulations.\\
In this contribution, we have introduced another way to build such processes, with a general form (Eq.~\ref{eq:New_Xt}) that requires regularizations. We have shown that specific regularizations yield existing processes, and we propose a new one, which has the benefits of relying on an infinite combination of Ornstein-Uhlenbeck processes. Characteristic time-scales of those Ornstein-Uhlenbeck are covering the inertial range, between Kolmogorov time-scale and the Integral time-scale. Each of them represents a specific turbulence structure, this corresponds to a continuous cascade model where no arbitrary time-scale is needed. \\
We have presented only essential ingredients of stochastic calculus for the purpose of presenting the new framework form a physical perspective but the details of the mathematical foundations can be found in a companion paper \cite{Goudenege2021}.
 A discrete version of this process is proposed, based on a selection of few specific modes, corresponding to representative characteristic time-scales. The quadrature of the infinite sum is therefore a finite sum of Ornstein-Uhlenbeck processes, logarithmically distributed in the inertial range. This corresponds to a discrete cascade model. \\
Beside the simplicity of simulation, this model has the benefit to be very adaptable and can be envisioned to be useful for future perspectives: dissipation along trajectory of solid inertial particles is not logarithmic anymore but the model can actually fit any auto-correlation function. \\
Thanks to the versatility of the process, application to LES can also be considered, with different regularizing functions, where the cut-off could be based on the subgrid time-scale for instance. 

\section*{Acknowledgements}
This work was supported by grants from Region Ile-de-France DIM MATHINNOV.
Support from the French Agence Nationale de la Recherche in the MIMETYC
project (grant ANR-17-CE22-0003) is also acknowledged.

\appendix

\section{Moments of the dissipation and coarse-grained dissipation}
\label{sec:Moments_dissipation_calculations}
The GMC formalism gives  $\varphi(t) = \mean{\varphi} \exp(\chi_t)$, with $\chi_t= \sqrt{\mu^\ell} X_t - \dfrac{\mu^\ell}{2} \mean{X_t^2}$.
It is immediate that the variance of $\chi_t$ can be expressed with the variance of $X_t$: $\sigma_\chi^2 = \mu^\ell \sigma_X^2$.\\
$\chi_t$ is a Gaussian variable, with moments generating function equal to: 
\begin{equation}
M_\chi(p) = \mean{\exp(p \chi_t)} = \exp \left( p \mu_\chi + \frac{1}{2}p^2 \sigma_\chi^2 \right)
\label{eq:MGF_Gaussian}
\end{equation}
Using $\mu_\chi = -\frac{1}{2}\sigma_\chi^2 $, this simplifies to: 
\begin{equation}
\mean{\varphi^p} = \mean{\varphi}^p \exp \left( p(p-1) \dfrac{\sigma_\chi^2}{2}  \right)
\label{eq:varphi_moments}
\end{equation}

Moments of the coarse-grained dissipation can also be derived as a function of the auto-correlation of $X_t$:
\begin{equation}
\begin{array}{ll}
(\varphi_\tau)^p&= \dfrac{1}{\tau^p} \displaystyle \int_{[t,t+\tau]^p} \prod\limits_{i=1}^p \varphi(s_i) \diff s_i \\
&= \dfrac{\mean{\varphi}^p}{\tau^p} \displaystyle \int_{[t,t+\tau]^p}  \exp \left(\sum\limits_{i=1}^p \sqrt{\mu^\ell} X_{s_i} - p \frac{\mu^\ell}{2}\sigma_X^2 \right)  \prod\limits_{i=1}^p \diff s_i  							
\end{array}
\end{equation}
Using the well-known identity $\mean{\exp(g)} = \exp(\frac{\mean{g^2}}{2})$for any zero-average Gaussian variable $g$, we have:
\begin{equation}
\begin{array}{ll}
\mean{\varphi_\tau^p} &=  \dfrac{\mean{\varphi}^p}{\tau^p} \mathrm{e}^{- p \frac{\mu^\ell}{2}\sigma_X^2} \displaystyle \int_{[t,t+\tau]^p}   \exp \Big( \frac{1}{2} \sum\limits_{i,j=1}^p  \mu^\ell \mean{X_{s_i} X_{s_j}} \Big)  \prod\limits_{i=1}^p \diff s_i  \\
&=  \dfrac{\mean{\varphi}^p}{\tau^p} \mathrm{e}^{- p \frac{\mu^\ell}{2}\sigma_X^2} \displaystyle \int_{[t,t+\tau]^p} \exp \Big( \sum\limits_{i<j} \mu^\ell \mean{X_{s_i} X_{s_j}} + p\frac{\mu^\ell}{2} \sigma_X^2 \Big)  \prod\limits_{i=1}^p \diff s_i  \\
&=  \mean{\varphi}^p  \displaystyle \int_{[0,1]^p} \exp \Big(  \mu^\ell \sum\limits_{i<j} \mean{X_{\tau s_i} X_{\tau s_j}} \Big)  \prod\limits_{i=1}^p \diff s_i  
\end{array}
\label{eq:varphi_tau_moments}
\end{equation}
%\begin{equation}
%\begin{array}{ll}
%\mean{\varphi_\tau^p} &=  \dfrac{\mean{\varphi}^p}{\tau^p} \mathrm{e}^{- p \frac{\mu^\ell}{2}\sigma_X^2} \displaystyle \int_{[t,t+\tau]^p}  \hspace*{-0.7cm} \exp \Big( \frac{1}{2} \sum\limits_{i,j=1}^p  \mu^\ell \mean{X_{s_i} X_{s_j}} \Big)  \prod\limits_{i=1}^p \diff s_i  \\
%&=  \dfrac{\mean{\varphi}^p}{\tau^p} \mathrm{e}^{- p \frac{\mu^\ell}{2}\sigma_X^2} \displaystyle \int_{[t,t+\tau]^p} \hspace*{-0.7cm} \exp \Big( \sum\limits_{i<j} \mu^\ell \mean{X_{s_i} X_{s_j}} \\
%& \displaystyle \hspace*{5cm} + p\frac{\mu^\ell}{2} \sigma_X^2 \Big)  \prod\limits_{i=1}^p \diff s_i  \\
%&=  \mean{\varphi}^p  \displaystyle \int_{[0,1]^p} \hspace*{-0.5cm} \exp \Big(  \mu^\ell \sum\limits_{i<j} \mean{X_{\tau s_i} X_{\tau s_j}} \Big)  \prod\limits_{i=1}^p \diff s_i  
%\end{array}
%\label{eq:varphi_tau_moments}
%\end{equation}
The last line results from a change of variables and the stationarity of the processes. 

\begin{comment}
\section{Laplace transform of the fractional Brownian motion}
\label{sec:Laplace_fBm}

\cite{Mandelbrot1968} suggested to use the Weyl integral to define the fBm claiming that the Liouville fractional integral formulation puts too great importance on the origin.They defined: 
\begin{equation}
W^H(t) = W^H(0) + \dfrac{1}{\Gamma(H+1/2)} \left( \int_{-\infty}^0 \left[ (t-s)^{H-1/2} - (-s)^{H-1/2} \right] \diff W_s 
+ \int_0^t (t-s)^{H-1/2} \diff W_s \right)
\end{equation}

Applying Laplace transformation on each term we get: 
\begin{equation}
\begin{array}{ll}
W^H(t) &= \displaystyle W^H(0) + \dfrac{1}{\Gamma(H+1/2)} \left( \int_{-\infty}^0 \left[ \int_0^\infty \mathrm{e}^{-x(t-s)} k(x) - \int_0^\infty \mathrm{e}^{xs} k(x) \right] \diff W_s 
+ \int_0^t \int_0^\infty \mathrm{e}^{-x(t-s)} k(x)\diff W_s \right) \diff x \\
&= \displaystyle W^H(0) + \dfrac{1}{\Gamma(H+1/2)} \int_0^\infty  \left( \left[ \mathrm{e}^{-xt} -1\right] \int_{-\infty}^0  \mathrm{e}^{xs}  \diff W_s 
+ \int_0^t \mathrm{e}^{-x(t-s)} \diff W_s \right) k(x) \diff x
\end{array}
\end{equation}

If we note $Y_0^x = \displaystyle \int_{-\infty}^0  \mathrm{e}^{xs}  \diff W_s$, and $Y_t^x = \displaystyle Y_0^x \mathrm{e}^{-xt} + \int_0^t  \mathrm{e}^{-x(t-s)} \diff W_s$

\begin{equation}
W^H(t) = \displaystyle W^H(0) + \dfrac{1}{\Gamma(H+1/2)} \int_0^\infty Y_t^x  k(x) \diff x
\end{equation}
\end{comment}

\section{Two-points correlation of Ornstein-Uhlenbeck correlated processes}
\label{sec:N_points_correlation_OU}
%Let us write $Y_t^x = \displaystyle Y_0^x \mathrm{e}^{-xt} + \int_0^t \mathrm{e}^{-(t-x) x} \diff W_s$. 
%\begin{equation}
%\begin{array}{ll}
%\mean{ Y_t^{x_i} Y_{t+\tau}^{x_j}} &=  \mean{Y_0^{x_i} \mathrm{e}^{-x_i t} Y_0^{x_j} \mathrm{e}^{-x_j (t+\tau)} } + \mean{\displaystyle \int_{0}^{t}\mathrm{e}^{-x_i(t-s)} \diff W_s   \int_{0}^{t+\tau}\mathrm{e}^{-x_j(t+\tau-s)} \diff W_s} \\
%
%&=   \mean{Y_0^{x_i} Y_0^{x_j}} \mathrm{e}^{-(x_i +x_j)t} \mathrm{e}^{-x_j \tau} + \mathrm{e}^{-(x_i+x_j)t}  \mathrm{e}^{-x_j\tau}   \displaystyle \int_{0}^{t}\mathrm{e}^{(x_i+x_j)s}\diff s \\
%
%& =  \dfrac{\mathrm{e}^{-x_j \tau}}{x_i + x_j} + \mathrm{e}^{-x_j \tau}\mathrm{e}^{-(x_i + x_j) t} \left(  \mean{Y_0^{x_i} Y_0^{x_j}} - \dfrac{1}{x_i + x_j}\right) 
%\end{array}
%\label{eq:2points_corrOU}
%\end{equation}
%
%Taking $Y_0^x = \displaystyle \int_{-\infty}^0 \mathrm{e}^{xs} \diff W_s$ ensures the stationarity because $ \mean{Y_0^{x_i} Y_0^{x_j}} = \dfrac{1}{x_i + x_j}$. 

Let us write $Y_t^x = \displaystyle  \int_{-\infty}^t \mathrm{e}^{-(t-x) x} \diff W_s$. For any $x_i ,x_j \in [0,+\infty[$ and $t>0$ and $\tau>0$, we have: 
\begin{equation}
\begin{array}{ll}
\mean{ Y_t^{x_i} Y_{t+\tau}^{x_j}} &=   \mean{\displaystyle \int_{-\infty}^{t}\mathrm{e}^{-x_i(t-s)} \diff W_s   \int_{-\infty}^{t+\tau}\mathrm{e}^{-x_j(t+\tau-s)} \diff W_s} \\
&=  \mathrm{e}^{-(x_i+x_j)t}  \mathrm{e}^{-x_j\tau}   \displaystyle \int_{-\infty}^{t}\mathrm{e}^{(x_i+x_j)s}\diff s \\
& =  \dfrac{\mathrm{e}^{-x_j \tau}}{x_i + x_j} 
\end{array}
\label{eq:2points_corrOU}
\end{equation}

%\begin{widetext}

\section{Variance and auto-correlation of $X_t^\infty$}
\label{sec:Calculations_Xt}
We calculate the auto-correlation of the process 
\begin{equation}
X_t^\infty = \displaystyle \int_0^\infty  Y_t^x \dfrac{1}{\sqrt{x}}   \left( g_{T_L}(x) - g_{\tau_\eta}(x) \right) \diff x 
\end{equation}

We denote $I = \mean{X_t^\infty X_{t+\tau}^\infty}$. We can consider, without loss of generality: $\mathrm{e}^{-\tau y} = 1 - g_\tau(y)$.  
\begin{equation}
\begin{array}{ll}
I &=  \displaystyle \int_0^\infty \int_0^\infty  \dfrac{\mathrm{e}^{-\tau y}}{(x+y)\sqrt{xy}}  \left( g_{T_L}(x) - g_{\tau_\eta}(x) \right)  \left( g_{T_L}(y) - g_{\tau_\eta}(y) \right)
\diff x \diff y \\

&=  \displaystyle \int_0^\infty \int_0^\infty  \dfrac{1}{(x+y)\sqrt{xy}}  \left( g_{T_L}(x) - g_{\tau_\eta}(x) \right)  \left( g_{T_L}(y) - g_{\tau}(y) \right)
\diff x \diff y \\

&=4  \displaystyle \int_0^{\pi/2} \int_{r=0}^\infty \dfrac{\left( g_\tau(r^2) - g_{T_L}(r^2) \right)  \left( g_{\tau_\eta/\tan^2 \theta}(r^2) - g_{T_L/\tan^2 \theta}(r^2) \right)}{r} \diff r \diff \theta 
\end{array}
\label{eq:autocorrelation_process}
\end{equation}
by using the transformation $(x,y) = (r^2 \cos^2 \theta,r^2 \sin^2 \theta)$ whose Jacobian is 
\begin{equation}
- 2r\cos(\theta)^{2} \times 2r^{2}\cos(\theta)\sin(\theta) - 2r^{2}\cos(\theta)\sin(\theta)\times 2r\sin(\theta)^{2}
=-4r^{3} \cos(\theta)\sin(\theta).
\end{equation}
The integral $I$ can be splitted in 5 parts according to the value of $\theta$. We introduce the functions $A,B,C,D,E$ all defined by the product $\left( g_\tau(r^2) - g_{T_L}(r^2) \right)  \left( g_{\tau_\eta/\tan^2 \theta}(r^2) - g_{T_L/\tan^2 \theta}(r^2) \right)$ but for different ranges of $\theta$.
\begin{equation}
\begin{array}{ll}
I/4 
&= \displaystyle \int_0^{\tan^{-1} \sqrt{\tau_\eta/T_L }} \int_{r=0}^\infty \dfrac{A(r^2)}{r} \diff r \diff \theta 
		+ \int_{\tan^{-1} \sqrt{\tau_\eta/T_L }}^{\tan^{-1} \sqrt{\tau_\eta/\tau }} \int_{r=0}^\infty \dfrac{B(r^2)}{r} \diff r \diff \theta \\
		&+\displaystyle\int_{\tan^{-1} \sqrt{\tau_\eta/\tau }}^{\pi/4} \int_{r=0}^\infty \dfrac{C(r^2)}{r} \diff r \diff \theta 
		+ \int_{\pi/4}^{\tan^{-1} \sqrt{T_L/\tau }} \int_{r=0}^\infty \dfrac{D(r^2)}{r} \diff r \diff \theta \\
		&+ \displaystyle \int_{\tan^{-1} \sqrt{T_L/\tau }}^{\pi/2} \int_{r=0}^\infty \dfrac{E(r^2)}{r} \diff r \diff \theta 
\end{array}
\end{equation}

\begin{figure}
\centering
 \includegraphics[height=5cm]{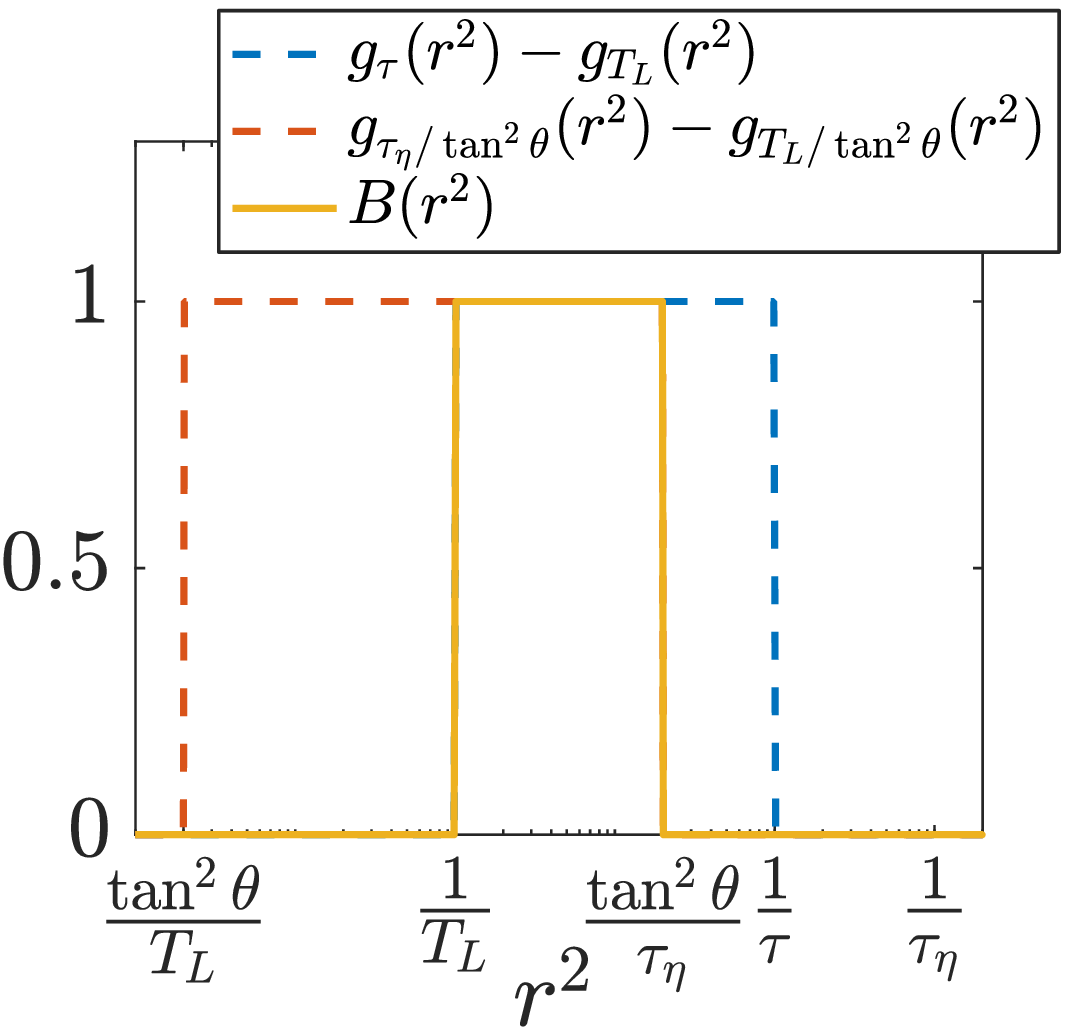}
  \includegraphics[height=5cm]{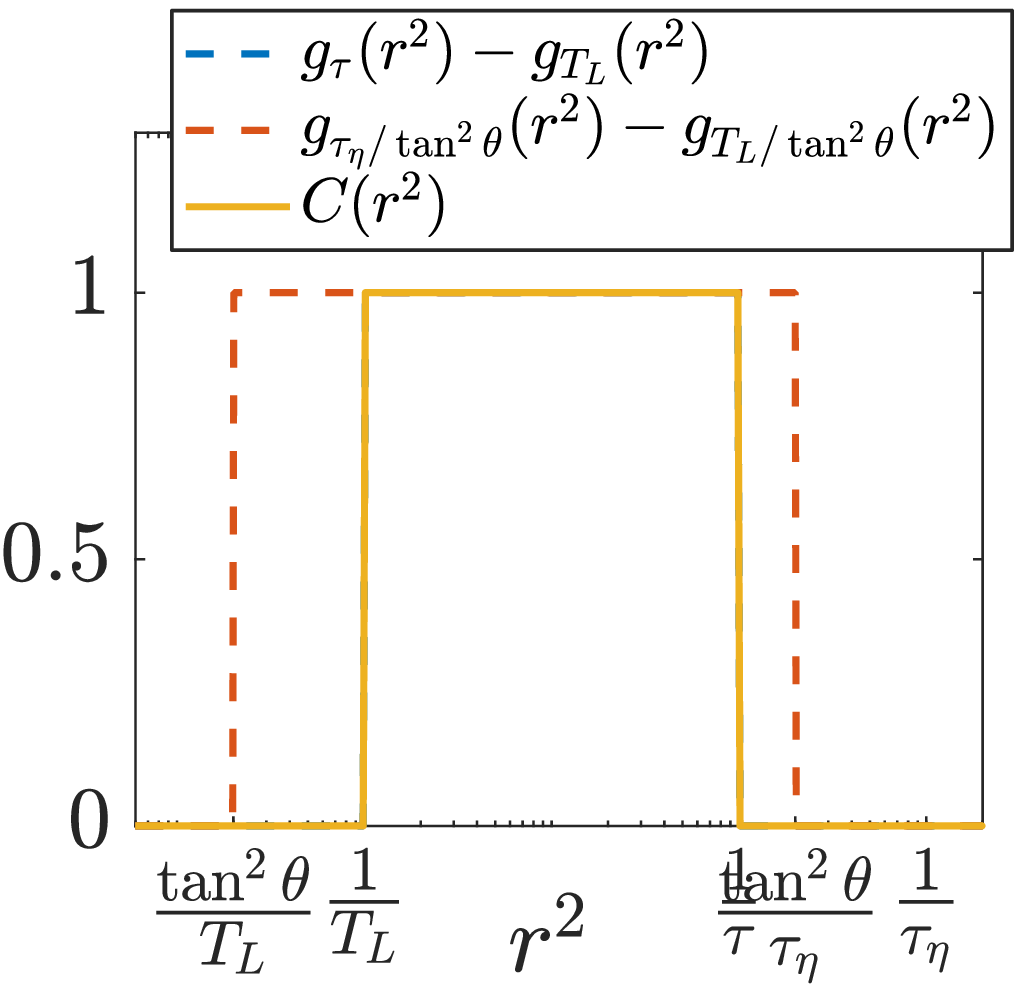} 
  \includegraphics[height=5cm]{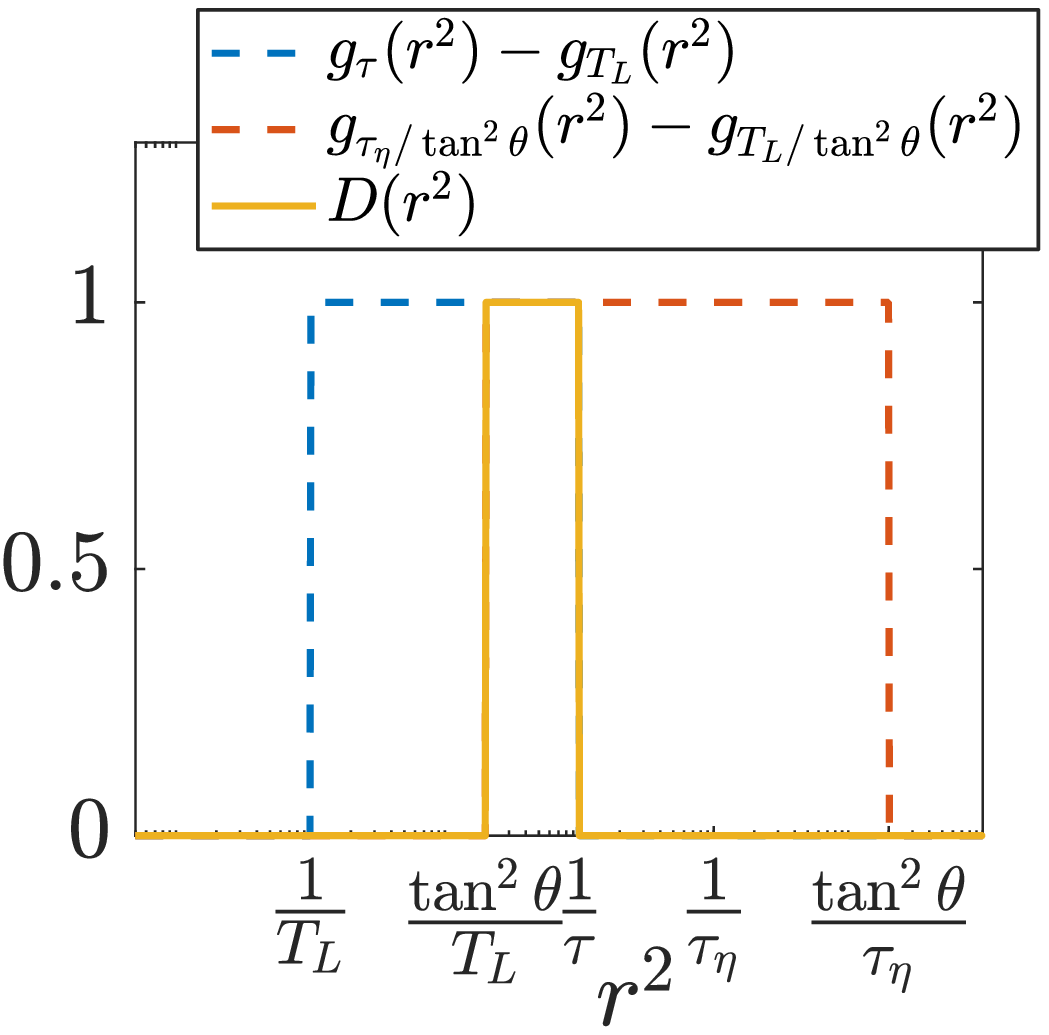}
\caption{Functions $B$, $C$, $D$ plotted for three examples of $\theta$}
\label{Cutting}
\end{figure}

To help the reader visualize the products of the regularized $g$-functions, we show in Fig.~\ref{Cutting} the schemes for the resulting product of the $g$-functions. 
The first and last integral are equal to zero because the “door" functions do not have any superposition. 
We can see from Fig.~\ref{Cutting} that $B$, $C$ and $D$ simplifies to:
\begin{equation}
\begin{array}{ll}
B(r^2) &= g_{T_L}(r^2) -  g_{\tau_\eta / \tan^2 \theta}(r^2)\\
C(r^2) &=  g_{T_L}(r^2) - g_{\tau}(r^2)\\
D(r^2) &= g_{T_L/\tan^2 \theta}(r^2) -g_{\tau}(r^2)  \\
\end{array}
\end{equation}
We use the property of the regularizing functions $g_\alpha$:
\begin{equation}
\displaystyle \int_0^\infty \dfrac{g_{\tau_1}(r^2) - g_{\tau_2}(r^2) }{r} \diff r \approx \int_{\sqrt{\tau_1}}^{\sqrt{\tau_2}} \dfrac{1}{r} \diff r = \dfrac{1}{2} \log \dfrac{\tau_2}{\tau_1}
\end{equation}
\begin{small}
\begin{equation}
\begin{array}{ll}
I/4 &= \displaystyle \int_{\tan^{-1} \sqrt{\tau_\eta/T_L }}^{\tan^{-1} \sqrt{\tau_\eta/\tau }}\dfrac{1}{2} \log \left( \dfrac{T_L \tan^2 \theta}{\tau_\eta} \right) \diff \theta 
+\displaystyle\int_{\tan^{-1}  \sqrt{\tau_\eta/\tau }}^{\pi/4} \dfrac{1}{2} \log \left( \dfrac{T_L}{\tau} \right) \diff \theta 
+ \int_{\pi/4}^{\tan^{-1} \sqrt{T_L/\tau }} \dfrac{1}{2} \log \left( \dfrac{T_L}{\tau \tan^2 \theta} \right) \diff \theta \\
		
&= \dfrac{1}{2} \log \dfrac{T_L}{\tau_\eta} \left( \tan^{-1} \hspace*{-0.2cm}\sqrt{\dfrac{\tau_\eta}{\tau}} - \tan^{-1}\hspace*{-0.2cm} \sqrt{\dfrac{\tau_\eta}{T_L}} \right)	
+ \dfrac{1}{2} \log \dfrac{T_L}{\tau} \left( \dfrac{\pi}{4} - \tan^{-1} \hspace*{-0.2cm}\sqrt{\dfrac{\tau_\eta}{\tau}} \right)
+ \dfrac{1}{2} \log \dfrac{T_L}{\tau} \left( \tan^{-1}\hspace*{-0.2cm} \sqrt{\dfrac{T_L}{\tau}} -  \dfrac{\pi}{4} \right)	\\
&+ \displaystyle \int_{\tan^{-1}\hspace*{-0.2cm} \sqrt{\tau_\eta/T_L }}^{\tan^{-1}\hspace*{-0.2cm} \sqrt{\tau_\eta/\tau }}  \log( \tan \theta) \diff \theta
\hspace*{7cm} - \displaystyle \int_{\pi/4}^{\tan^{-1} \sqrt{T_L/\tau }}  \log( \tan \theta) \diff \theta		\\
I 	&\underset{\tau_\eta \ll \tau \ll T_L}{\sim} \pi \log \left( \dfrac{T_L}{\tau} \right)  + 4 \displaystyle  \int_{0}^{\pi/4} \log ( \tan \theta) \diff \theta  
\end{array}
\end{equation}
\end{small}

The variance can be deduced from this calculation:
\begin{equation}
\mean{(X_t^\infty)^2} =  \displaystyle \int_0^\infty \int_0^\infty  \dfrac{1}{(x+y)\sqrt{xy}}  \left( g_{T_L}(x) - g_{\tau_\eta}(x) \right)  \left( g_{T_L}(y) - g_{\tau_\eta}(y) \right)
\diff x \diff y 
\end{equation}
We remark that this expression is similar to the one obtained in Eq.\ref{eq:autocorrelation_process} where $\tau$ is replaced by $\tau_\eta$. Therefore, we obtain: 
\begin{equation}
\mean{(X_t^\infty)^2} = \pi \log \left( \dfrac{T_L}{\tau_\eta} \right)  + 8 \displaystyle  \int_{0}^{\pi/4} \log ( \tan \theta) \diff \theta  
\end{equation}
% \end{widetext}

\bibliography{library}% Produces the bibliography via BibTeX.

\end{document}